%% file: main.tex
\documentclass[12pt]{article}
\usepackage[utf8]{inputenc}
\usepackage[T1]{fontenc}
\usepackage{lmodern}
\usepackage{microtype}
\usepackage{authblk}
\usepackage{amsmath}
\usepackage{amssymb}
\usepackage{graphicx}
\usepackage{subcaption}
\usepackage{booktabs}
\usepackage[margin=1in]{geometry}
\usepackage[authoryear,round]{natbib}
\usepackage{xcolor}
\usepackage{url}
\usepackage{tikz}
\usetikzlibrary{arrows.meta, positioning, calc}

\title{Domain-Adapted Deep Optical Flow for Multiphase Velocimetry in Unsteady Cloud Cavitation}
\author[1]{Naga Nitish Chamala}
\author[1]{Sanjay Vasanth Kethanur Balasubramaniam}
\author[1]{Olivier Coutier-Delgosha}
\affil[1]{Kevin T. Crofton Department of Aerospace and Ocean Engineering, Virginia Tech, Blacksburg, VA 24061, USA}
\affil[ ]{Corresponding author: chnaganitish@gmail.com}

\date{}

\begin{document}

\maketitle

\begin{abstract}
Time-resolved velocimetry inside cloud cavitation remains a fundamental measurement problem. Particle image velocimetry (PIV) requires uniform tracer seeding and undistorted particle images, and cavitating flow violates both, because tracers do not penetrate vapor structures and refractive index mismatch across deforming interfaces produces spurious vectors. We present a domain-adapted deep optical flow framework for two-phase velocimetry, built on a Recurrent All-Pairs Field Transforms (RAFT) architecture fine-tuned on synthetic image pairs rendered from multiphase OpenFOAM simulations (RAFT-CloudCav). It is evaluated against a pre-trained generalist RAFT model, Farneb\"ack and OpenPIV on $130{,}000$~frames/s recordings of a Venturi test section at $30$ and $48$~L/min. As no experimental ground truth exists, validation uses three quantities that require none, namely the integrated flow rate against a turbine meter, the liquid-phase velocity divergence, which mass conservation requires to vanish, and the station-to-station spread of the flow rate. The four methods agree within $2\%$ in the liquid core and integrate to $41.6\text{--}43.2$~L/min against a metered $48.0 \pm 0.5$~L/min. The shortfall of $10\text{--}13\%$ is common to all four and so lies in the measurement rather than in any one estimator, and part of it is traced to the coarse imaging configuration that spans the channel. RAFT-CloudCav gives the most consistent flow rate between stations, varying by $1.8\%$ across four streamwise stations against $4.2\text{--}4.5\%$ for the other methods. At $48$~L/min the learned estimators reduce the liquid-phase divergence three- to seventeenfold relative to the classical ones, with the pre-trained model lowest, and OpenPIV under-reads the cavitating near-wall shear layer by up to $38\%$. Every method resolves the re-entrant jet in the instantaneous field, the learned ones with less directional scatter. In the non-cavitating control the boundary layer is captured better as the imaging is refined. At $49$~$\mu$m/pixel only three grid points fall inside a $0.66$~mm boundary layer and the peak wall-normal gradient is under-predicted nearly sixfold, while $13$ and $19$~$\mu$m/pixel agree with each other within $6\%$. The mean field is recovered without tracer particles to within $13\%$, at about five times the per-frame uncertainty. Uncertainties are estimated \emph{a posteriori} from the signed image residual and propagated conservatively.
\end{abstract}

\vspace{0.5em}

\section{Introduction}
Cavitation, the formation, growth, and collapse of vapor cavities in a liquid, governs performance and erosion in hydraulic machinery, marine propulsors, and fuel injection systems \citep{brennen1995}. Among its regimes, cloud cavitation is the most violent and the least tractable. A sheet cavity grows from a low-pressure region, an adverse pressure gradient at the cavity closure drives a liquid re-entrant jet upstream beneath it, the jet severs the sheet, and the detached cloud is convected downstream and collapses \citep{callenaere2001, kawanami_mechanism_1997}. The shedding is periodic, and the mechanism that terminates each cycle, whether re-entrant jet cutting or a propagating condensation shock, depends on operating point \citep{hatzissawidis_transition_2025}. In a Venturi geometry the process is further shaped by interfacial instability at the cavity shear layer \citep{podbevsek_kelvin_2021} and by the structure of the near-wall reverse flow itself \citep{gawandalkar_structure_2022}.

Quantifying these mechanisms requires velocity measurements inside the two-phase region, at a temporal resolution matched to the shedding. PIV is the established tool for flow-field diagnostics \citep{thielicke_pivlab_2014, scharnowski_particle_2020}, and it has been applied to cavitating flows around hydrofoils \citep{foeth_time_2006, zhang_piv_2024, simanto_comparative_2025}. But its assumptions are structurally at odds with cavitation. Cross-correlation requires a locally uniform seeding density, which vapor regions do not provide. Tracer particles are excluded from the cavity interior, and their transport is itself altered near interfaces \citep{poelma_measurement_2020}. The refractive index mismatch across a deforming liquid--vapor boundary distorts and displaces particle images, casts shadows, and produces correlation peaks that do not correspond to fluid motion \citep{poelma_measurement_2020, chen_two-phase_2024}. Outlier detection and iterative image deformation mitigate but do not remove the problem \citep{westerweel_universal_2005, scarano2002}, because the underlying signal is absent rather than merely noisy.

Optical flow offers a different premise. Rather than correlating interrogation windows, it estimates a dense per-pixel displacement from the spatial and temporal intensity structure of the image pair \citep{horn_determining_1981}. It does not require discrete particles, only a texture that persists between frames, and a cavitating flow supplies abundant texture in the form of vapor structures, micro-bubbles, and density gradients. Classical variational and polynomial formulations have been adapted to fluids for this reason \citep{liu_fluid_2008, corpetti_dense_2002, farneback_two-frame_2003}, extended with wavelet bases for time-resolved measurement \citep{derian_wavelets_2013, schmidt_high-resolution_2019}, and the field has been surveyed recently \citep{jassal_review_2025}. Their limitation is the regularization that makes them well-posed. A smoothness prior penalizes exactly the steep velocity gradients that a cavitation shear layer contains, so the method tends to under-predict the quantity of most interest.

Deep optical flow architectures learn the regularizer from data instead of imposing it. RAFT \citep{vedaldi_raft_2020} builds an all-pairs correlation volume and refines a flow field iteratively with a recurrent update operator, and lighter architectures trade accuracy for cost \citep{hui_liteflownet_2018}. Applied to fluids, learned estimators have been demonstrated for PIV data \citep{cai_particle_2020, lagemann2021deep}, for bubbly flows \citep{choi_bubble_2022}, for bubble segmentation \citep{kim_deep_2021}, and more broadly across fluid mechanics \citep{brunton_machine_2020}. A systematic study of RAFT-PIV on synthetic and experimental data \citep{lagemann_generalization_2022} showed that its performance depends on the image and lighting conditions it is applied to. The premise of the present work is that the training data must represent the conditions of the target experiment. In cavitation specifically, deep networks have so far been applied to image segmentation rather than velocimetry, as in the modified U-Net that separates sheet from shed cloud in high-speed cloud-cavitation photography \citep{hatzissawidis_deep_2026}. Velocity inside the two-phase region has instead been reached by ultra-fast X-ray imaging \citep{khlifa_fast_2017}, which resolves the flow within the cavity but requires a synchrotron beamline.

The advantages come with costs that deserve to be stated plainly. A learned estimator inherits the statistics of its training set, and generalist optical flow benchmarks are dominated by rigid-body motion, planar translation, and rendered characters, a domain far from deforming liquid--vapor interfaces. That gap must be closed by adapting the estimator to the target domain \citep{zhou_unsupervised_2023}. Supervised fine-tuning, the route taken here, requires ground-truth displacement fields that an experiment cannot supply, and synthetic rendering from simulation is the usual answer \citep{mendes_piv-image-generator_2020, kalagotla_new_2024}, and datasets have been assembled for related two-phase problems \citep{hassan_bubbleml_2023}. A network also returns no intrinsic estimate of its own reliability. Unlike cross-correlation, which exposes a signal-to-noise ratio per vector, a learned flow field is equally confident where it is right and where it is wrong. And its learned prior remains a prior, which can silently smooth a gradient it has not seen.

This paper presents a velocimetry framework for cloud cavitation built on a fine-tuned RAFT model (RAFT-CloudCav), and evaluates it against a pre-trained RAFT model, the Farneb\"ack algorithm, and OpenPIV on high-speed Venturi recordings. Because no ground truth exists for the experiment, the evaluation is built on three measurements that need none. These are the integrated volumetric flow rate against an independent turbine meter, the velocity divergence in the liquid phase against mass conservation, and the consistency of the flow rate between streamwise stations. The instantaneous fields are also examined for the re-entrant jet, whose existence and direction are known from the physics. A per-frame uncertainty is estimated and propagated to every time-averaged quantity, and used to establish an operating envelope in imaging resolution and seeding. The measurement-science contribution is a velocimetry technique that returns velocities inside the two-phase region, together with an \emph{a posteriori} uncertainty estimate and a validation route that requires no reference velocity field. The measurement pipeline is released as an open-source package.

\section{Methodology}

\subsection{Experimental Setup and High-Speed Imaging}
\label{sec:setup}
\begin{figure}[p]
    \centering
    \begin{subfigure}{\linewidth}
        \centering
        \includegraphics[width=0.82\linewidth]{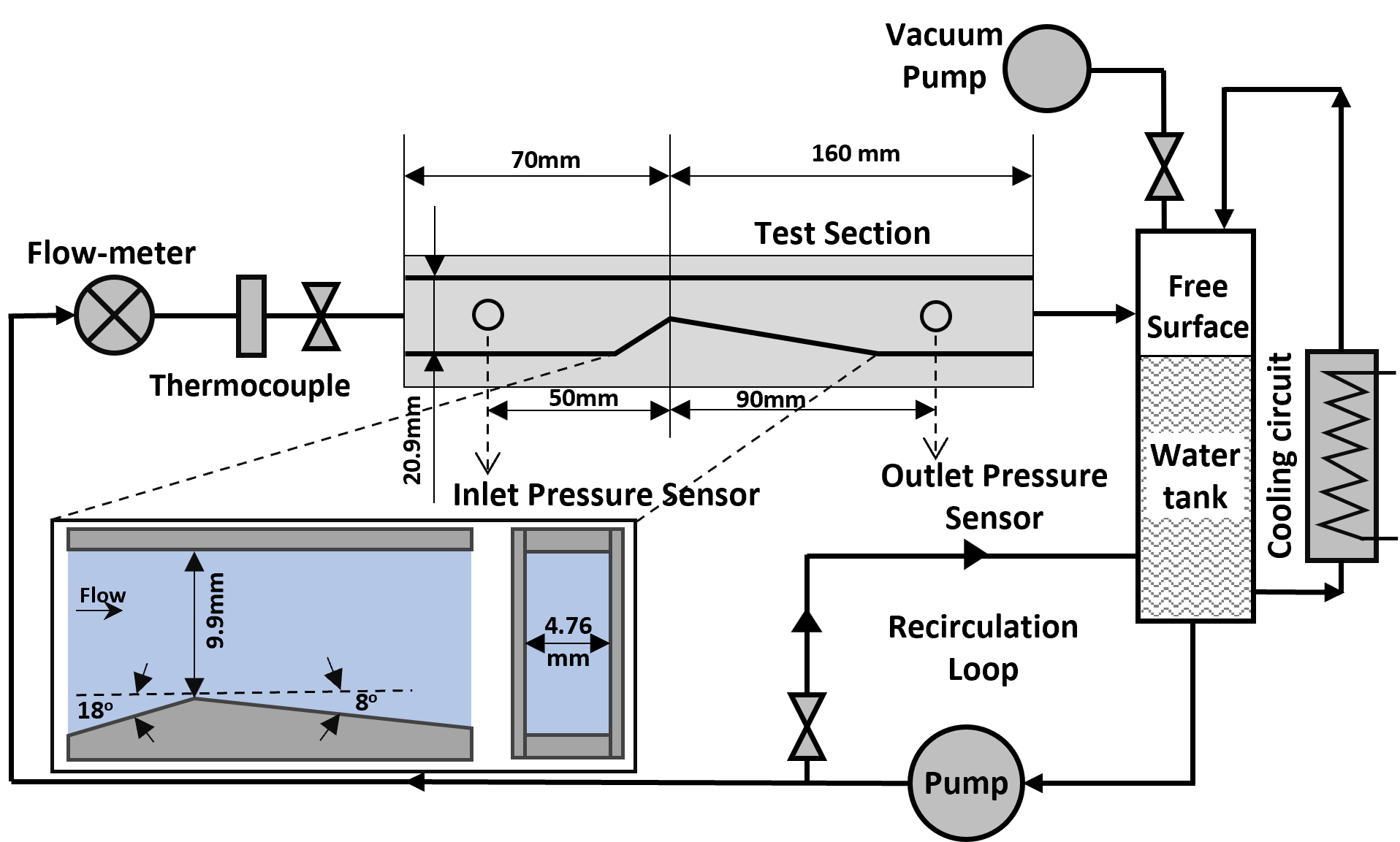}
        \caption{Closed-loop cavitation tunnel}
        \label{fig:setup_loop}
    \end{subfigure}

    \vspace{4mm}
    \begin{subfigure}{\linewidth}
        \centering
        \resizebox{!}{0.27\textheight}{\input{tikz/camera_lighting}}
        \caption{Camera and illumination in (i) top view and (ii) side view}
        \label{fig:setup_optics}
    \end{subfigure}
    \caption{Experimental setup. (a) The closed-loop cavitation tunnel. Water is driven through the Venturi test section by the recirculation pump and returns to the reservoir. A turbine flow meter records the volumetric flow rate, a thermocouple the water temperature, and pressure sensors the static pressure upstream and downstream of the throat. A cooling circuit in the reservoir holds the temperature and a vacuum pump acting on the free surface of the reservoir sets the loop pressure. (b) The imaging arrangement. In the top view (i) the camera views the test section through one acrylic side wall, with its optical axis normal to that wall, so the recorded field is a projection through the $4.76$~mm channel depth, and the continuous LED sits directly behind the test section on the opposite side. The side view (ii), looking downstream with the camera on the near side, shows that the LED is raised above the far side of the test section and tilted $45^\circ$ to the optical axis, so its beam enters the channel obliquely and the camera collects light scattered by structure inside the cavity rather than a backlit silhouette. The effect of this arrangement on the images is shown in figure~\ref{fig:sample-frames}.}
    \label{fig:Experimental_setup}
\end{figure}

Experiments were conducted in the closed-loop cavitation tunnel used by \citet{zhang_venturi_2020} (figure~\ref{fig:Experimental_setup}), with a Venturi-type test section of the class used extensively for unsteady cloud cavitation \citep{coutier-delgosha_analysis_2007, barre2009, brunhart_investigation_2020}. The channel has a constant width $w = 4.76\text{ mm}$, an inlet height of $20.9\text{ mm}$, and a throat height $H = 9.9\text{ mm}$. The convergent and divergent section angles are $18^\circ$ and $8^\circ$ respectively. Flow was driven by a Salmson circulation pump under variable-frequency control and metered by a turbine flow meter, which read $48.0 \pm 0.5$~L/min at the cavitating condition. The water temperature was held at $20$~$^\circ$C by a cooling circuit in the reservoir and monitored with a thermocouple upstream of the test section. Static pressure was measured by two piezoresistive pressure sensors (Keller 10L series, calibrated over $0$--$3$~bar with a full-scale uncertainty of $0.25\%$) located $50$~mm upstream and $90$~mm downstream of the throat (figure~\ref{fig:setup_loop}), and the reservoir pressure was set by a vacuum pump acting on the free surface of the water tank. The dissolved gas content was not measured.

\begin{figure}[h!]
    \centering
    \includegraphics[width=0.98\linewidth]{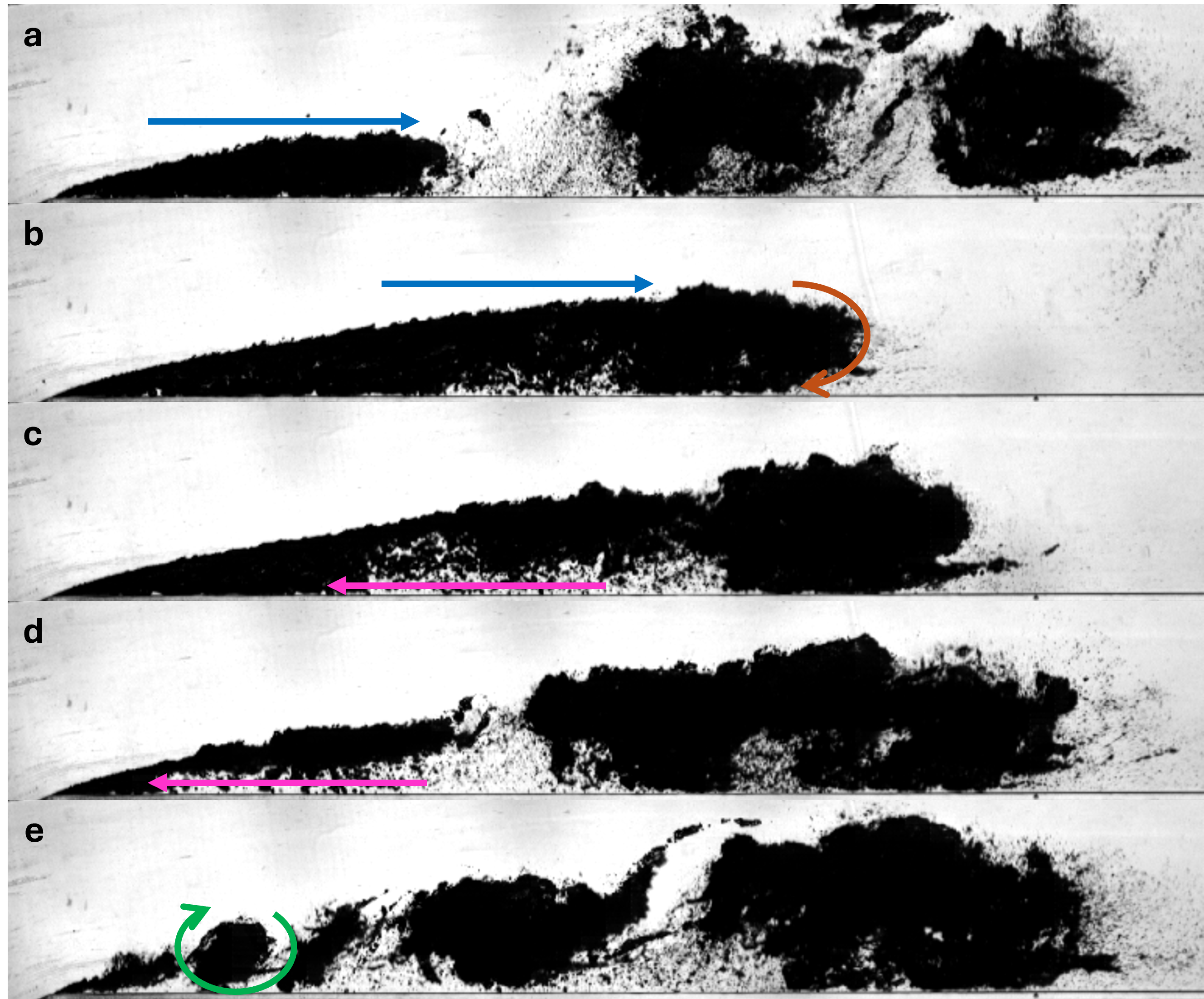}
    \caption{The cloud cavitation cycle, recorded in the Venturi test section at $48\text{ L/min}$. Flow is from left to right and the sheet cavity attaches at the throat, near the left edge of each panel. (a) The sheet cavity grows downstream of the throat as the pressure drops along the divergent section, and the straight blue arrow marks the downstream growth of the cavity front. (b) The cavity reaches its maximum length and an adverse pressure gradient at the closure turns liquid underneath it. The curved orange arrow marks that turning. (c) The straight pink arrow marks the liquid re-entrant jet running upstream along the wall beneath the cavity. (d) The jet reaches the throat and the sheet breaks into a cloud of bubbles, with the pink arrow again marking the jet. (e) The shed cloud rolls up under Kelvin--Helmholtz instability, marked by the circular green arrow, and convects downstream to collapse. The recording is at approximately $50~\mu$m/pixel and is shown to illustrate the cycle, so no scale bar is drawn.}
    \label{fig:cloudcavitation}
\end{figure}

Two operating points are reported. The first, $30$~L/min, is a non-cavitating condition that serves as a control, and the second, $48$~L/min, produces unsteady cloud cavitation (figure~\ref{fig:cloudcavitation}). The bulk velocity at the throat, $U_b = Q/(wH)$, is $10.6$ and $17.0$~m/s respectively, giving throat Reynolds numbers $\mathrm{Re} = U_b H/\nu$ of $1.05 \times 10^5$ and $1.68 \times 10^5$ for water at $20~^\circ$C, $\nu = 1.0 \times 10^{-6}$~m$^2$/s. The cavitation number, based on the outlet pressure, $\sigma_c = (p_\mathrm{out} - p_v)/(\tfrac{1}{2}\rho U_b^2)$, is $1.64$ at $30$~L/min and $0.60$ at $48$~L/min.

Recordings were acquired through transparent acrylic side walls with a Phantom TMX 6410 high-speed camera at $130{,}000$~frames/s and a $1~\mu$s exposure, which limits motion blur to $20~\mu$m at $20$~m/s, or $1.5$~pixels in the finest configuration. The sensor window was $1280 \times 384$~pixels at this frame rate, read from the recordings themselves, with a $100$~mm lens. Each record comprises $13{,}624$ frames, giving $13{,}623$ consecutive frame pairs and approximately $105\text{ ms}$ of continuous observation, which spans several tens of shedding cycles. Three seeded imaging configurations were used for the resolution study, at spatial calibrations of $13.0$, $19.2$, and $49.2$~$\mu$m/pixel, referred to below as the fine, medium, and coarse configurations. Each calibration was obtained with a dot-pattern calibration plate imaged against the front and the back outer walls of the test section, giving scale factors $C_f$ and $C_b$. The scale factor at the plane of focus inside the channel, $C_p = 2C_fC_b/[(C_f + C_b) + (w/W)(C_b - C_f)]$, was then interpolated from these, where $w = 4.76$~mm is the channel width and $W$ is the distance between the two outer walls. The plane of focus is the front face of the channel. The procedure is set out in Appendix~\ref{app:calibration}. Unseeded recordings were acquired separately at $20.5$ and $56.7$~$\mu$m/pixel, the former close enough to the medium configuration to support the comparison of Section~\ref{sec:seeding}. The camera sensor window has a fixed number of pixels, so a finer calibration gives a smaller field of view, and table~\ref{tab:configs} lists the resulting fields of view. In the wall-normal direction the fine configuration covers $4.37$~mm and the coarse configuration $9.83$~mm. The channel height grows downstream of the throat as $H(x) = H + x\tan 8^\circ$, from $9.9$~mm at the throat to $10.60$~mm at $x = 5$~mm, so the coarse configuration spans $99\%$ of the channel height at the throat and $93\%$ of it at the most downstream station used for integral quantities. Only the coarse configuration is used for those quantities.

\begin{table}[htbp]
    \centering
    \caption{Imaging configurations. Calibration, sensor window and processed region are read from the recordings and the HDF5 metadata. Frames are cropped to the fluid domain before tracking, so the wall-normal extent quoted here is that of the processed region, not of the full sensor height, while the streamwise extent is the full sensor width. Fields of view are the processed region in pixels multiplied by the calibration. $H(x) = 9.9 + x\tan 8^\circ$~mm, so the channel height is $9.9$~mm at the throat and $10.60$~mm at $x = 5$~mm. Peak displacement is the largest time-averaged speed in the processed region, converted to pixels per frame, at each flow rate. The synthetic training pairs span about $2.6$--$3.8$~px/frame (Section~\ref{sec:synthetic}).}
    \label{tab:configs}
    \small
    \resizebox{\linewidth}{!}{%
    \begin{tabular}{lccccccccc}
        \toprule
         & Calibration & & Sensor window & Processed region & Field of view & \multicolumn{2}{c}{Channel height covered} & \multicolumn{2}{c}{Peak displ. (px/frame)} \\
        \cmidrule(lr){7-8}\cmidrule(lr){9-10}
        Configuration & ($\mu$m/pixel) & Seeding & (pixels) & $1280 \times n_y$ & $x \times y$ (mm) & at throat & at $x=5$~mm & 30 L/min & 48 L/min \\
        \midrule
        Fine         & 13.0 & seeded, $10~\mu$m & $1280 \times 384$ & $n_y = 336$ & $16.6 \times 4.37$  & 44\%  & 41\%  & 7.9 & 10.9 \\
        Medium       & 19.2 & seeded, $40~\mu$m & $1280 \times 384$ & $n_y = 320$ & $24.6 \times 6.14$  & 62\%  & 58\%  & 5.3 & 7.5 \\
        Coarse       & 49.2 & seeded, $40~\mu$m & $1280 \times 384$ & $n_y = 200$ & $62.9 \times 9.83$ & 99\% & 93\%  & 1.9 & 2.7 \\
        Unseeded, medium & 20.5 & none & $1280 \times 384$ & $n_y = 360$ & $26.2 \times 7.38$  & 75\%  & 70\%  & -- & 6.5 \\
        Unseeded, coarse$^{*}$ & 56.7 & none & $1280 \times 384$ & $n_y = 176$ & $72.6 \times 9.98$  & 101\% & 94\%  & -- & -- \\
        \bottomrule
    \end{tabular}}

    \raggedright\footnotesize $^{*}$Recorded but not used in this paper.
\end{table}

The images are recorded through the $4.76$~mm channel depth with volume illumination and no light sheet, so the measured field is a projection. It is weighted by the depth of field of the lens, which was not quantified, and averages the in-plane velocity of all trackable texture across the depth.

Both artificially seeded conditions and unseeded conditions, relying on natural micro-bubbles and fluid impurities, were recorded. The seeding was silver-coated hollow glass spheres of mean diameter $d_p = 10~\mu$m for the fine configuration and $d_p = 40~\mu$m for the medium and coarse configurations, whose larger fields of view need larger particle images, and nominal density $\rho_p \approx 1.4$~g\,cm$^{-3}$, typical of this particle type. The seeding concentration was not quantified. The geometric particle image diameters are about $0.8$, $2.1$ and $0.8$~pixels in the fine, medium and coarse configurations, so in the fine and coarse configurations the particles are imaged below one pixel and are tracked through their diffraction-broadened images.

With $\rho_p \approx 1400$~kg\,m$^{-3}$ and water at $\mu \approx 1.0 \times 10^{-3}$~Pa\,s, the particle response time $\tau_p = \rho_p d_p^2/(18\mu)$ is $7.8~\mu$s for the $10~\mu$m particles and $124~\mu$s for the $40~\mu$m particles. Against the bulk timescale at $48$~L/min, $H/U_b \approx 0.58$~ms, this gives Stokes numbers $\mathrm{Stk} \approx 0.013$ and $0.21$ respectively (at $30$~L/min, $H/U_b \approx 0.93$~ms and $\mathrm{Stk} \approx 0.008$ and $0.13$), so both follow the bulk flow, the larger particles less closely. The most demanding timescale in the measurement is the peak near-wall shear, $46.5$~(m/s)/mm at $30$~L/min, for which $1/(\partial U/\partial y) \approx 22~\mu$s. For the $10~\mu$m particles of the fine configuration, where this gradient was measured, $\mathrm{Stk} \approx 0.36$ and the tracers lag somewhat in the steepest near-wall shear. For the $40~\mu$m particles the same gradient gives $\mathrm{Stk} \approx 5.8$, so they do not follow the near-wall shear at its steepest, and in the medium and coarse configurations the near-wall velocity is carried by the motion of the particle images and of any other texture together rather than by faithful tracers. This is one reason the unseeded mean field remains informative. The gravitational settling velocity, $(\rho_p - \rho_f) g d_p^2/(18\mu)$, is $2 \times 10^{-5}$ and $3.5 \times 10^{-4}$~m/s for the two sizes, negligible against the bulk velocity in both cases. The free-stream nuclei size distribution was not measured.

The test section carried a visible mechanical vibration during the seeded runs, while the unseeded recordings were free of it. A rigid displacement of the section appears in the image as a whole-field translation, and its amplitude in \emph{pixels} scales inversely with the calibration, so the same physical vibration is about four times larger, in pixel terms, in the fine configuration than in the coarse one. Where it is a uniform translation it is real image motion, every estimator tracks it as velocity, and the image residual of Section~\ref{sec:uq} does not register it. Any part of it that is not a uniform in-plane translation, such as a rotation, flexing of the section or motion through the depth of field, changes the image between frames in a way the warp cannot reproduce, and that part does raise the residual. Vibration therefore enters the measurement both as a fluctuating bias in the velocity and as part of the reported uncertainty, and both grow with magnification. Its amplitude was not measured, although it could be recovered from the apparent motion of a static feature such as the wall edge.

\begin{figure}[h!]
    \centering
    \includegraphics[width=0.82\linewidth]{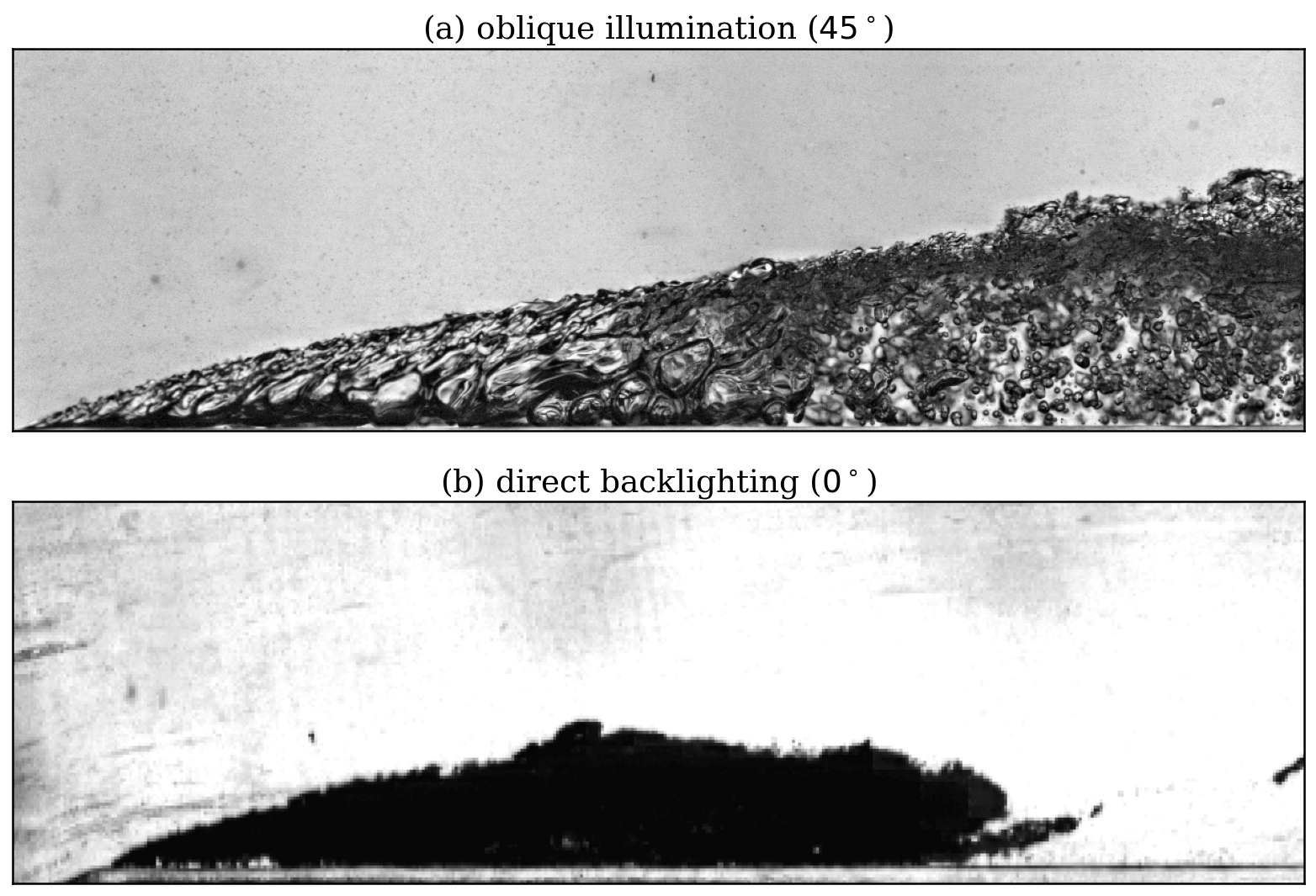}
    \caption{Influence of the lighting arrangement on the visibility of internal structure within vapor clouds, for the two arrangements sketched in figure~\ref{fig:setup_optics}. Oblique illumination (a) preserves the texture that optical flow tracks, resolving individual bubbles, the roughness of the interface, and structure within the cloud. Direct backlighting (b) reduces the same kind of cavity to a featureless silhouette, leaving the estimator nothing to match inside it. Flow is from left to right in both. The two recordings are at different magnifications, so the wider frame has been cropped to the aspect ratio of the other for display and the panels show comparable but not identical fields of view. Both recordings are at approximately $50~\mu$m/pixel and are shown for illustration.}
    \label{fig:sample-frames}
\end{figure}

A high-intensity continuous LED source was placed directly behind the test section, raised and tilted $45^\circ$ to the optical axis (figure~\ref{fig:setup_optics}), rather than as direct backlighting. As Figure~\ref{fig:sample-frames} shows, direct backlighting renders vapor structures as featureless silhouettes, whereas oblique illumination reveals internal gradients and textural variation within the vapor through light scattering. This distinction matters more for optical flow than for PIV, because the method tracks texture and an illumination that flattens the vapor into a silhouette removes the signal the estimator depends on.

\subsection{Velocimetry Algorithms}
\label{sec:algorithms}
\begin{figure}[p]
    \centering
    \resizebox{\linewidth}{!}{\input{tikz/raft_architecture}}
    \caption{The RAFT-CloudCav architecture and its training pipeline. A preprocessed image pair passes through a shared-weight feature encoder at $1/8$ resolution, while a context encoder sees $I_t$ alone. All-pairs correlation between the two feature maps builds a 4D cost volume, pooled into a four-level pyramid. A recurrent ConvGRU operator repeatedly looks up correlation values around the current flow estimate and emits an increment $\Delta\mathbf{u}_k$, and the converged estimate is upsampled to full resolution by a learned convex combination. RAFT-CloudCav and RAFT-Sintel share this architecture exactly and differ only in their weights. RAFT-CloudCav is fine-tuned on OpenFOAM-rendered pairs with known ground truth, whereas RAFT-Sintel is the generalist checkpoint used without adaptation. Downstream, every method feeds the same block-averaging and statistical reduction.}
    \label{fig:raft_arch}
\end{figure}
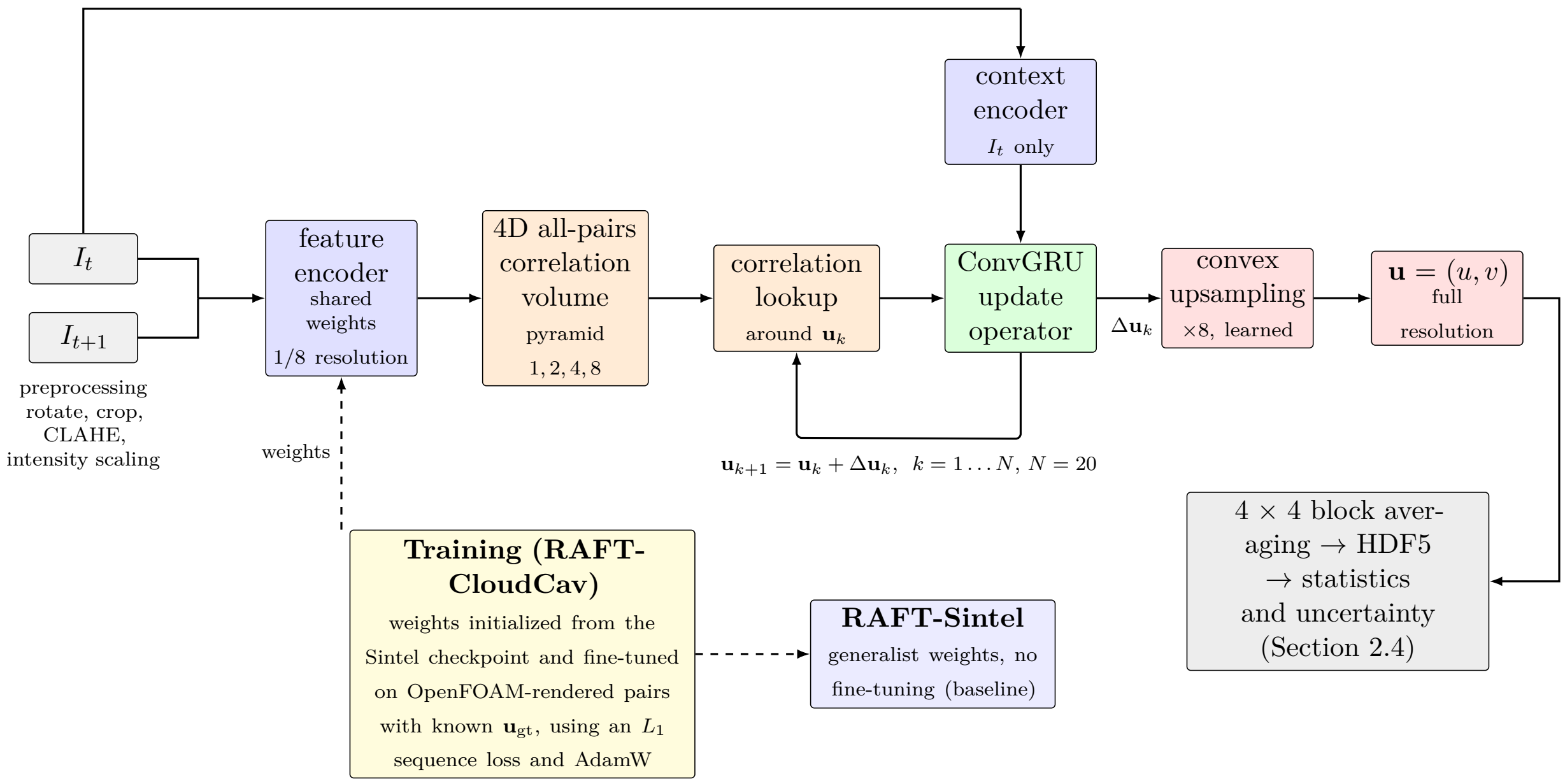

Four algorithms were applied to identical preprocessed image pairs. The two RAFT variants share the architecture sketched in figure~\ref{fig:raft_arch} and differ only in their weights.

\begin{enumerate}
    \item \textbf{Farneb\"ack} is classical dense optical flow by polynomial expansion \citep{farneback_two-frame_2003}, run via OpenCV with a pyramid scale of $0.5$, three levels, a window size of $15$, three iterations per level, a polynomial neighborhood of $5$ pixels and a Gaussian width of $1.2$. These are standard settings and were not tuned to the recordings.    \item \textbf{OpenPIV} is multi-pass cross-correlation with window deformation, using interrogation windows of $64 \rightarrow 32 \rightarrow 16 \rightarrow 8$ pixels at $50\%$ overlap, Gaussian sub-pixel peak fitting, and signal-to-noise validation, following standard practice \citep{scarano2002}.
    \item \textbf{RAFT-Sintel} is the same architecture with weights pre-trained on the generalist Sintel benchmark, included to isolate the contribution of domain adaptation from that of the architecture.    \item \textbf{RAFT-CloudCav} is the RAFT architecture \citep{vedaldi_raft_2020} fine-tuned on synthetic two-phase image pairs generated as described in Section~\ref{sec:synthetic}, using the AdamW optimizer, a variant of Adam with decoupled weight-decay regularization \citep{loshchilov_decoupled_2019}, at a learning rate of $2 \times 10^{-5}$ and weight decay $10^{-5}$, with an $L_1$ sequence loss over the recurrent updates weighted by $\gamma = 0.85$. Starting from the Sintel checkpoint, the network was fine-tuned for $3000$ steps at a batch size of $8$, about $33$ passes over the $723$ training pairs, on random $1288 \times 208$ pixel crops of the $1296 \times 216$ pixel pairs, on one GPU, with $12$ recurrent iterations during training. The crops leave at most $8$ pixels of random translation, so they act as little more than a jitter. At inference the recurrent operator is run for $20$ iterations.\end{enumerate}

All four were run through a common pipeline so that preprocessing, spatial averaging, and statistical reduction are identical across methods and only the estimator differs. Raw $10$-bit sensor counts were first scaled to the $0$--$255$ range in floating point rather than quantized. Frames were then rotated to level the divergent wall, cropped to the fluid domain, and contrast-enhanced with contrast-limited adaptive histogram equalization (CLAHE) \citep{zuiderveld_contrast_1994} at a clip limit of $2.0$ and an $8 \times 8$ tile grid, and its effect on a recorded frame is shown in Appendix~\ref{app:clahe}. The RAFT models receive these floating-point frames directly. The OpenCV Farneb\"ack implementation and OpenPIV require integer images, so their frames are rounded to $8$ bits. Dense flow fields were block-averaged onto a $4 \times 4$ pixel grid and written to chunked HDF5 with the calibration, capture rate, and processing parameters recorded alongside, so that every downstream quantity is traceable to the run that produced it.

\subsection{Multiphase CFD and Synthetic Image Generation}
\label{sec:synthetic}
Training data were generated from three-dimensional large-eddy simulations (LES) of the same Venturi geometry in OpenFOAM. The numerical framework follows the one validated for this Venturi by \citet{apte_investigation_2024,apte_numerical_2024}, which uses the \texttt{interPhaseChangeFoam} solver, a transient homogeneous mixture with volume-of-fluid interface capturing \citep{hirt_volume_1981}, and the Merkle transport-equation model for mass transfer between the phases \citep{merkle_computational_1998}. Those studies used detached-eddy simulation on meshes of up to $8.4 \times 10^6$ cells, whereas the present simulations use LES on the meshes given below. These simulations generated the CFD data used by the synthetic image generator described below. Grid convergence was established across coarse ($1.2 \times 10^5$ cells), medium ($4.5 \times 10^5$), and fine ($1.8 \times 10^6$) meshes for both velocity profiles and shedding frequency.

\begin{figure}[h!]
    \centering
    \begin{subfigure}{\linewidth}
        \centering
        \includegraphics[width=\linewidth]{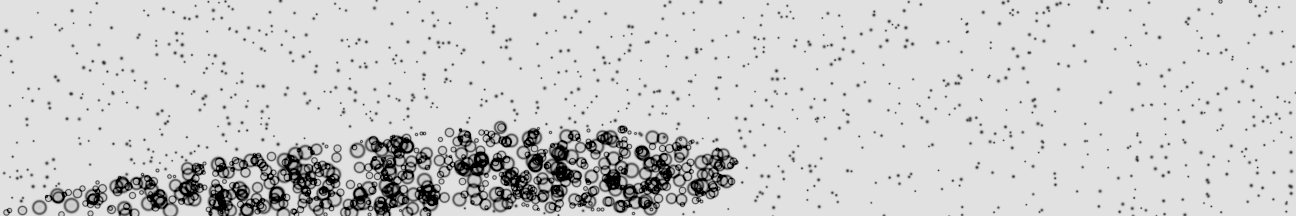}
        \caption{$I_t$}
    \end{subfigure}

    \vspace{0.2cm}
    \begin{subfigure}{\linewidth}
        \centering
        \includegraphics[width=\linewidth]{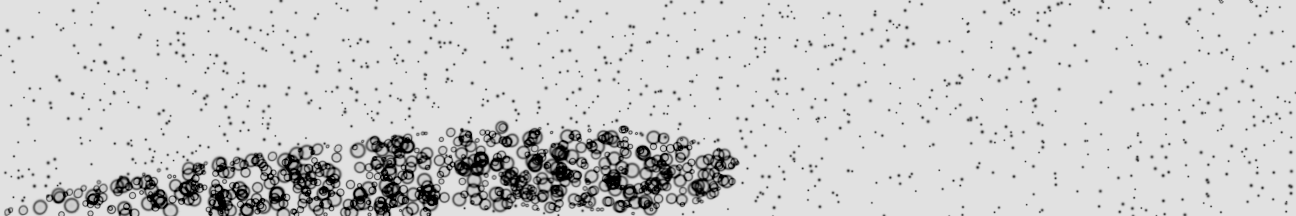}
        \caption{$I_{t+1}$}
    \end{subfigure}
    \caption{A synthetic image pair $(I_t, I_{t+1})$ rendered from the three-dimensional OpenFOAM void fraction and velocity fields on a quarter-plane of the channel, at $1296 \times 216$ pixels and the shortest of the three frame intervals, $\Delta t = 3.5~\mu$s. The pair is the last time step of the held-out test block. Flow is from left to right, and each frame spans $60$~mm streamwise and $10$~mm wall-normal.}
    \label{fig:synth_reference}
\end{figure}

Velocity and void-fraction fields from the fine-grid simulation, sampled on the mid-plane and on the two quarter-planes of the channel width, were coupled to a synthetic image renderer, following the approach established for PIV and optical flow benchmarking \citep{mendes_piv-image-generator_2020, kalagotla_new_2024}. Each saved time step was rotated by $8^\circ$ about the throat, into the same wall-aligned frame as the experimental images, cropped to $0 \le x \le 60$~mm and $0 \le y \le 10$~mm, and interpolated onto a $1296 \times 216$ pixel grid, giving square pixels of $46.3~\mu$m. Vapor was rendered as rim-lit bubbles, a bright center with a dark interface ring on a uniform background. Bubbles were placed in the cloud core (water volume fraction $\alpha < 0.99$) at $150$ per $10^4$ pixels with radii of $2$--$8$ pixels, and along the interface ($0.99 \le \alpha < 0.999$) at $600$ per $10^4$ pixels with radii of $1$--$2$ pixels. The liquid ($\alpha \ge 0.99$) was seeded with dark Gaussian particle images of radius $1$--$2$ pixels at $40$ per $10^4$ pixels, drawn at their exact sub-pixel positions so that the displacement in the images matches the label. The second frame was produced by warping the bubble field with the displacement $\mathbf{u}_{\text{gt}} = \mathbf{U}\,\Delta t/\Delta x$, solving for the source position of each pixel by fixed-point iteration so that the warped bubbles stay consistent with the label at large displacements, advecting each particle individually, and seeding new particles where liquid enters the frame. Each plane was given its own frame interval, $\Delta t = 7~\mu$s on the mid-plane, close to the $7.7~\mu$s frame interval of the experiment, and $3.5$ and $21~\mu$s on the two quarter-planes, so that the set covers slower and faster motion than the mid-plane on the same pixel grid. Each plane was rendered separately as a sharp image pair, labelled with the velocity of that plane. A rendered pair is shown in figure~\ref{fig:synth_reference}. The $307$ saved time steps were split in time. The first four fifths were used for training and the last fifth for testing, with five steps left out between the two blocks, giving $723$ training and $183$ test pairs, one per plane and time step. Because the ground truth is known by construction, these pairs provide the supervision that an experiment cannot. Two properties of this set bound what it can show. First, the test pairs come from later in the same simulation, so they measure generalization to flow fields the model has not seen in this Venturi, not to a new geometry or operating point. Second, at $46.3~\mu$m/pixel a velocity of $17$--$25$~m/s is $1.3$--$1.9$, $2.6$--$3.8$ and $7.8$--$11.4$~px/frame on the three planes, which covers the peak displacements of all three experimental configurations, $1.9$ to $10.9$~px/frame (table~\ref{tab:configs}). The particle and bubble images are $2$--$16$~pixels across, however, so the particle images of the experiment are not matched in size.

\subsection{Uncertainty Quantification}
\label{sec:uq}
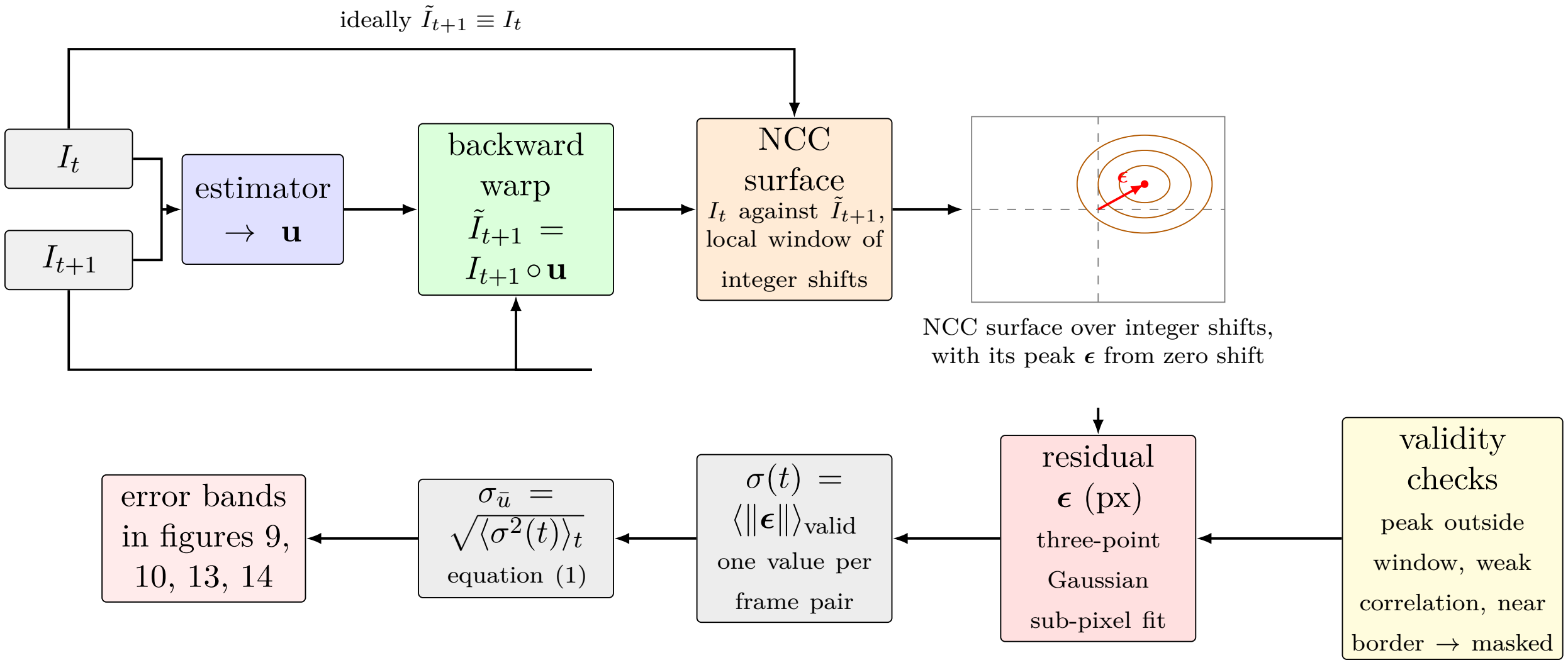
\begin{figure}[p]
    \centering
    \resizebox{\linewidth}{!}{\input{tikz/uq_concept}}
    \caption{The \emph{a posteriori} uncertainty estimate. The estimator predicts $\mathbf{u}$ from the image pair, and $I_{t+1}$ is then warped backward by that prediction, which in a perfect measurement would reproduce $I_t$ exactly. At each interrogation point the normalized cross-correlation between $I_t$ and the warped image is evaluated over a local window of integer shifts, and a three-point Gaussian fit locates its peak to sub-pixel precision. The offset of that peak from zero shift is the residual displacement $\boldsymbol{\epsilon}$, the displacement that the correlation indicates is still missing. The normalized cross-correlation is evaluated over an $8 \times 8$ pixel window and integer shifts of up to $\pm 4$ pixels. The residuals give a per-frame $\sigma(t)$, which combines through equation~(\ref{eq:sigmabar}) into the uncertainty bands shown on every profile.}
    \label{fig:uqconcept}
\end{figure}

A learned flow estimator does not report its own reliability, so its uncertainty is estimated \emph{a posteriori} from the image residual that remains after the predicted displacement has been applied. Cross-correlation methods expose several such measures, among them peak ratio, particle disparity, and correlation statistics, and their relative merits have been assessed collaboratively \citep{sciacchitano_collaborative_2015}. The approach used here is closest in spirit to the correlation-statistics formulation \citep{wieneke_piv_2015}, adapted to operate on the residual of an arbitrary dense estimator rather than on a correlation plane computed by the estimator itself. For each frame pair, the second image is backward-warped by the predicted flow, and in a perfect measurement the warped image would coincide with the first. A local window of integer pixel shifts is then searched for the peak of the normalized cross-correlation surface, located to sub-pixel precision by a three-point Gaussian fit. The offset between that peak and zero shift is the \emph{residual displacement error}, the additional displacement that correlation analysis indicates is still needed after warping by the predicted flow (figure~\ref{fig:uqconcept}).

Four properties of this quantity govern how it is used. First, it is the realized error of each individual frame pair, not the standard deviation of an error distribution, and part of it may be systematic, because a smoothness prior under-predicts the same gradient in the same place on every frame. Second, it is meaningful only where the correlation surface is informative. No points are masked in the statistics reported here. Where the correlation peak sits at the edge of the $\pm 4$ pixel search, the three-point fit is taken about the nearest interior shift, so the residual is limited to $\pm 4$ pixels and a poorly informed point contributes a large residual rather than being dropped. Within $12$ pixels of the frame edge, the search half-width plus the window, the shifted image is padded by reflection, so the three vector rows nearest each edge are evaluated against partly synthesized image content. Third, the residual is recorded with its sign, so at each point its time mean $b$, the part that recurs on every frame, can be separated from its fluctuation $\sigma_{\text{rand}}$ about that mean. Fourth, successive frames of an unsteady flow are correlated in time, so a time average contains fewer independent samples than frames.

The two limits for the uncertainty of a mean are far apart. If the residual were random in time it would average down as
$1/\sqrt{N_{\text{eff}}}$, with $N_{\text{eff}} = N/(2\tau_{\text{int}})$ set by
the integral timescale of the velocity autocorrelation \citep{benedict_towards_1996}.
For the fine configuration at $48$~L/min the median integral timescale across the profile at $x = 1.5$~mm is $\tau_{\text{int}} \approx 2.3$~frames, so the $13{,}623$ frame pairs supply $N_{\text{eff}} \approx 2900$ independent samples and would reduce the random part by a factor of about $54$. At $30$~L/min the velocity fluctuations decorrelate within one frame at most points, and $N_{\text{eff}}$ reaches its upper limit $N$. If instead
the residual is systematic, averaging removes none of it.

The signed residual shows which limit applies to the residual itself. At $x = 1.5$~mm in the fine configuration at $30$~L/min, RAFT-CloudCav returns a residual whose time mean is $|b| \approx 0.05$~m/s against a root-mean-square of $1.35$~m/s (medians across the profile), so almost all of it fluctuates from frame to frame. The measured variance then bounds how much of that fluctuation can be velocity error, since $\mathrm{Var}(u_{\text{meas}}) = \mathrm{Var}(u_{\text{true}})
+ \mathrm{Var}(\epsilon_{\text{rand}})$ with both terms non-negative, so
$\mathrm{Var}(\epsilon_{\text{rand}}) \le \langle u'u' \rangle$. At the same station $\langle u'u' \rangle = 0.91\text{ m}^2/\text{s}^2$ against a residual variance of $1.83\text{ m}^2/\text{s}^2$. A random velocity error as large as the residual is therefore ruled out, since its variance would be twice the measured one, and removing it from $\langle u'u' \rangle$ would leave a negative Reynolds stress. The residual therefore overstates the velocity error. Image changes that the warp cannot reproduce even with the correct flow, such as deforming interfaces and particles moving through the illuminated depth, and the noise floor of the sub-pixel fit all contribute to it. The external checks agree, since the fine and medium profiles agree to $0.57$~m/s although their residuals are $1.35$ and $0.66$~m/s (Section~\ref{sec:spatial}). How much of the residual is velocity error cannot be fixed without a coverage test against ground truth (Section~\ref{sec:limitations}).

The bands reported here nevertheless treat the whole residual as systematic, as an upper bound, which gives
\begin{equation}
    \label{eq:sigmabar}
    \sigma_{\bar{u}} = \sqrt{\left\langle \sigma^2(t) \right\rangle_t},
\end{equation}
where $\langle\cdot\rangle_t$ denotes the average over all frame pairs at that point. This is the conservative limit, since averaging can only reduce an uncertainty, and with it $\sigma_{\bar{u}} = \sqrt{b^2 + \sigma^2_{\text{rand}}}$ equals the instantaneous precision. The opposite limit, $\sqrt{b^2 + \sigma^2_{\text{rand}}/N_{\text{eff}}}$, in which the random part averages down, is $0.06$~m/s at the station above. The bands shown on every profile are therefore an upper bound on the image-matching error of the mean, and comparisons that they do not resolve may still be real.

This quantity covers image matching alone. It excludes the calibration uncertainty in $\mu$m/pixel, which enters velocity linearly, out-of-plane motion through the $4.76\text{ mm}$ channel depth, test-section vibration, and the systematic component of the estimator's own smoothing prior beyond what the residual captures. It also excludes the sampling uncertainty of the mean from a finite number of shedding cycles. Collecting every
contribution, the total uncertainty of a time-averaged velocity is
\begin{equation}
    \sigma^2_{U,\text{total}} = \sigma^2_{\text{match,sys}}
      + \frac{\sigma^2_{\text{match,rand}}}{N_{\text{eff}}}
      + \sigma^2_{\text{cal}} + \sigma^2_{\text{timing}}
      + \sigma^2_{\text{oop}} + \sigma^2_{\text{vib}} + \sigma^2_{\text{prior}},
\end{equation}
of which only the image-matching terms are quantified here, and those as an upper bound. Table~\ref{tab:budget}
lists each source and how it is treated.

\begin{table}[htbp]
    \centering
    \caption{Uncertainty budget for a time-averaged velocity. Only the image-matching terms are quantified, as an upper bound, and the remainder are identified but not evaluated.}
    \label{tab:budget}
    \small
    \begin{tabular}{p{0.26\linewidth}p{0.40\linewidth}p{0.26\linewidth}}
        \toprule
        Source & Treatment in this paper & Magnitude \\
        \midrule
        Image-matching residual, systematic & Quantified. The whole residual is conservatively treated as systematic, so $\sigma_{\bar{u}} = \sigma_{\text{match}}$ & $0.64$--$1.35$~m/s as profile medians and up to $4.84$~m/s inside the boundary layer, at $x = 1.5$~mm, seeded, $30$~L/min (Section~\ref{sec:spatial}) \\
        Image-matching residual, random & Separated from the systematic part by the signed residual and bounded by the variance argument. Not averaged down in the bands & Time mean $0.05$~m/s against a root-mean-square of $1.35$~m/s, and random velocity-error variance at most $50\%$ of the residual variance, for RAFT-CloudCav at $x = 1.5$~mm, fine configuration, $30$~L/min \\
        Test-section vibration & The uniform translation is tracked as real motion and biases the velocity. The remainder raises the residual and is included in it. Seeded runs only & Not quantified separately \\
        Spatial calibration ($\mu$m/pixel) & Not included. Enters velocity linearly. The scale varies through the channel depth between the front- and back-wall calibrations & Not quantified \\
        Frame timing & Not included & Negligible (camera clock) \\
        Out-of-plane motion and depth integration & Not included. Discussed qualitatively in Section~\ref{sec:limitations} & Not quantified \\
        Learned-prior smoothing undetected by the residual & Not included & Not quantified \\
        \bottomrule
    \end{tabular}
\end{table}

\subsection{Ground-Truth-Free Validation Metrics}
\label{sec:metrics}
No reference velocity field exists for the experiment, so two physical constraints are used in its place. Throughout, $x$ is the streamwise distance downstream of the throat, so that $x = 0$ at the throat, and $y$ is the wall-normal distance measured from the divergent (bottom) wall, increasing into the flow. The images are rotated so that the divergent wall is horizontal, and the velocity components $(u, v)$ are taken parallel and normal to that wall. At the throat the near-wall flow still carries the inclination of the $18^\circ$ convergent wall, which lies at $26^\circ$ to this frame, so $v$ is not zero at $x = 0$ and decays downstream as the flow turns to follow the divergent wall.

\textbf{Mass conservation.} In a single-phase incompressible region continuity requires $\nabla \cdot \mathbf{U} = 0$, and any measured departure is error. The metric is evaluated only within the liquid phase. Inside a vapor cavity the velocity divergence does not vanish, because vapor is produced and condensed there, and including vapor would penalize whichever method tracks the cavity most responsively. For the liquid, a uniform average across a channel of constant width with impermeable side walls satisfies two-dimensional continuity exactly, even in a three-dimensional flow. The projected measurement departs from that uniform average because it is weighted by the depth of field. The liquid mask is obtained by applying Otsu's threshold to the block-averaged brightness of each frame. The vapor cloud is the darker phase under this illumination (figure~\ref{fig:vectors}, Appendix~\ref{app:clahe}), so blocks above the threshold are kept as liquid, and gradients are not taken across the mask boundary. The metric is computed by central differences on the $4 \times 4$ pixel vector grid of the fine configuration, over $200$ randomly sampled frame pairs. Since $\nabla \cdot \mathbf{U}$ carries units of inverse time, a method that reports systematically slower flow would otherwise score well while being more wrong. Results are therefore reported as $|\nabla \cdot \mathbf{U}|\,H/U_{\text{ref}}$, with $U_{\text{ref}}$ the $95$th-percentile liquid velocity of that same measurement.

\textbf{Integrated flow rate.} The volumetric flow rate $Q = w \int U \, \mathrm{d}y$ across the channel is compared against the turbine meter. Two details determine whether the comparison is fair. Only the coarse configuration spans almost the full channel, so for the others part of the integral is extrapolation and is reported as such. And the divergent section opens at $8^\circ$, so the integration domain must follow the geometry as $H(x) = H + x\tan 8^\circ$. Integrating every station over the throat height instead truncates a growing fraction of the channel and manufactures an apparent mass loss with downstream distance. Where the flow is liquid across the whole height, mass conservation makes $Q$ independent of streamwise station, and the spread of $Q$ across stations is a second consistency check obtained from the same data. At $48$~L/min the sheet cavity starts at the throat, so the stations downstream of it cut through vapor. There the integral includes the velocity of the visible vapor structure, and $\int u\,\mathrm{d}y$ of the mixture is not conserved where vapor forms or condenses. The station spread is therefore a test of consistency between stations rather than a strict mass balance, and the throat value is the one compared with the meter. The station spread is defined as $(Q_{\max} - Q_{\min})/\bar{Q}$ over the four stations $x = 0$, $1.5$, $3$ and $5$~mm. Where the channel height exceeds the field of view, the integrand is closed linearly to zero at the domain boundary rather than truncated, and the fraction of the domain actually measured is reported with each value.

\subsection{Use of AI Tools}
\label{sec:ai}
A large language model, Claude (Anthropic), was used in preparing this manuscript, and its use is reported here. It was used to edit and restructure the text and to check every reference against its DOI record. It was also used to draft the post-processing and plotting scripts and to check the reported statistics against the archived data, including the extraction of the line profiles and the liquid-phase divergence analysis shown in figures~\ref{fig:div48} and \ref{fig:div30}. The explanatory schematics in figures~\ref{fig:setup_optics}, \ref{fig:raft_arch}, \ref{fig:uqconcept} and \ref{fig:calibration} are vector drawings written in TikZ code. The language model drafted the code from the authors' descriptions and sketches, and the authors edited and checked it. No image-generation model was used, and these schematics contain no measured data. The experiments, the measured and simulated data, the trained models, and the interpretation and conclusions are the authors' own. The authors reviewed and validated all AI-assisted output and take full responsibility for the content of the manuscript.

\section{Results and Discussion}

\subsection{Comparison on Synthetic Data}
\begin{figure}[p]
    \centering
    \begin{subfigure}{\linewidth}
        \centering
        \includegraphics[width=\linewidth]{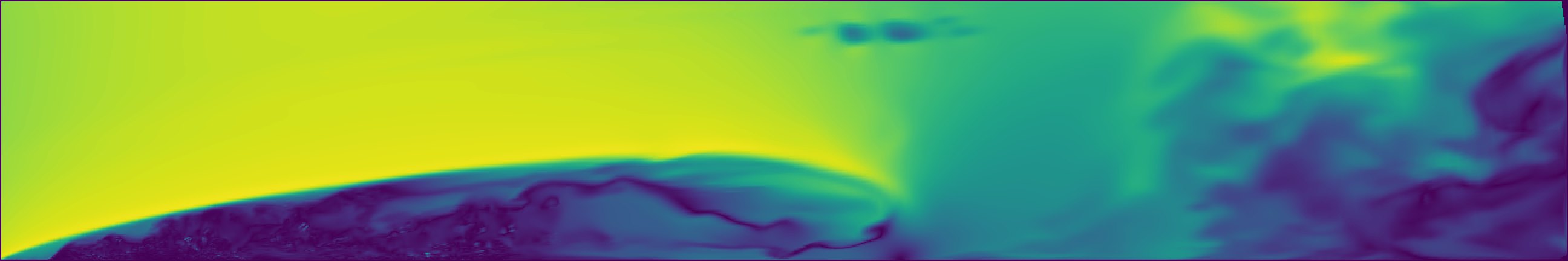}
        \caption{Ground truth from CFD}
        \label{fig:truVel}
    \end{subfigure}

    \vspace{0.1cm}
    \begin{subfigure}{\linewidth}
        \centering
        \includegraphics[width=\linewidth]{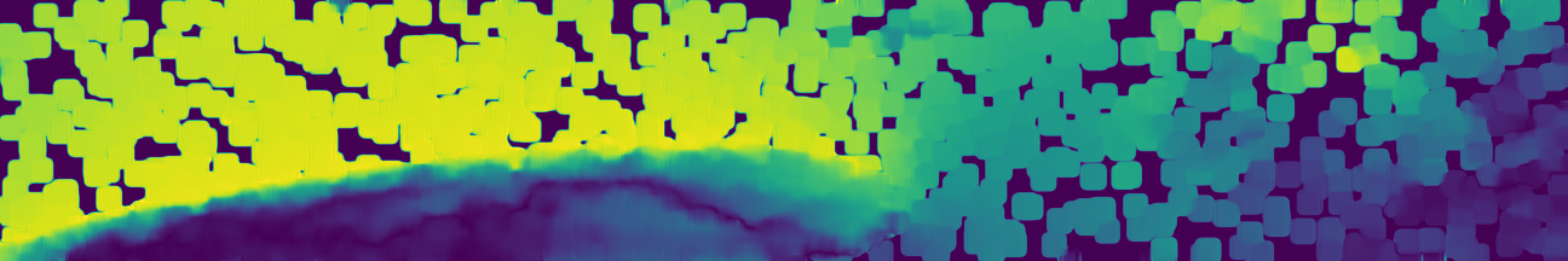}
        \caption{Farneb\"ack}
        \label{fig:fbVel}
    \end{subfigure}

    \vspace{0.1cm}
    \begin{subfigure}{\linewidth}
        \centering
        \includegraphics[width=\linewidth]{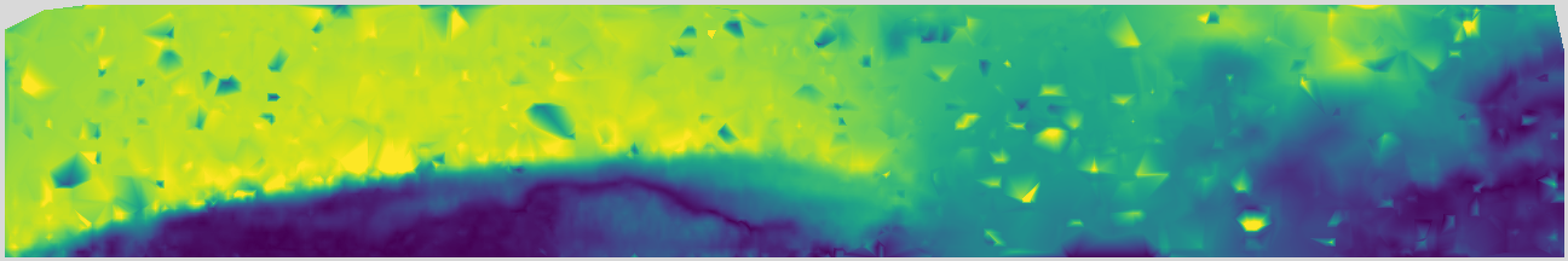}
        \caption{OpenPIV}
        \label{fig:pivVel}
    \end{subfigure}

    \vspace{0.1cm}
    \begin{subfigure}{\linewidth}
        \centering
        \includegraphics[width=\linewidth]{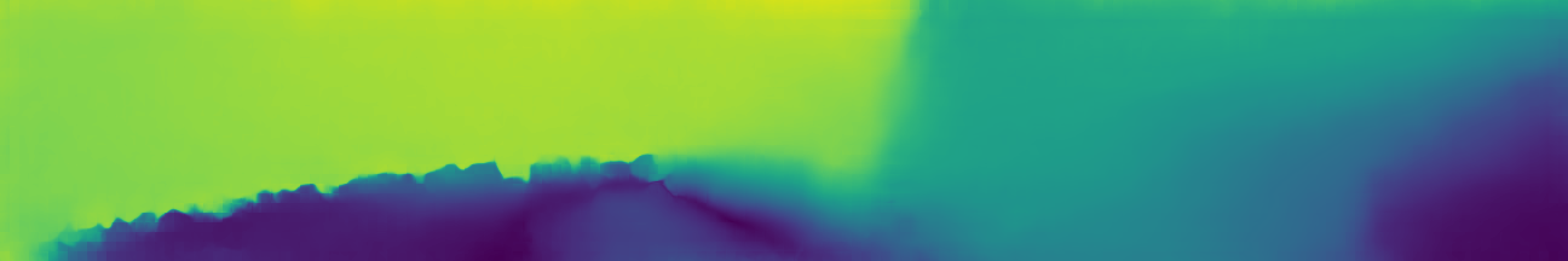}
        \caption{RAFT-Sintel (pre-trained)}
        \label{fig:rsVel}
    \end{subfigure}

    \vspace{0.1cm}
    \begin{subfigure}{\linewidth}
        \centering
        \includegraphics[width=\linewidth]{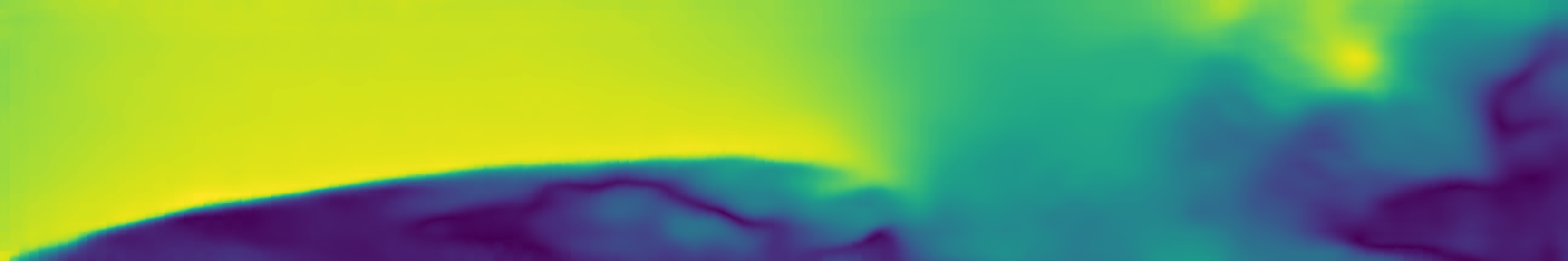}
        \caption{RAFT-CloudCav (fine-tuned)}
        \label{fig:rcVel}
    \end{subfigure}

    \vspace{0.2cm}
    \includegraphics[width=0.5\textwidth]{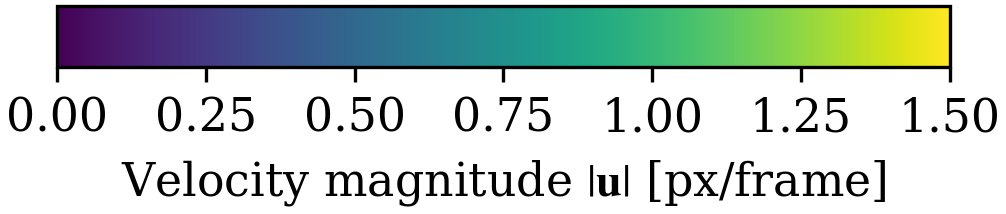}
    \caption{Velocity magnitude contours from each algorithm against synthetic ground truth for the test pair of figure~\ref{fig:synth_reference}, ordered from classical to learned after the ground truth. All panels share the color scale shown below them, the velocity magnitude $|\mathbf{u}|$ in px/frame. Gray marks the frame edge where OpenPIV returns no vectors. Flow is from left to right.}
    \label{fig:velociComparison}
\end{figure}

On the held-out test pair of figure~\ref{fig:synth_reference}, the four estimators differ most around the cavity (figure~\ref{fig:velociComparison}). Farneb\"ack returns a continuous field inside the bubbly cavity, where the texture is dense, but in the seeded liquid it reports motion only around individual particles and almost none between them. OpenPIV recovers the free stream and the cavity boundary, with small faceted patches where rejected vectors were dropped and the gaps interpolated. RAFT-Sintel returns a smooth field but a ragged cavity interface, and it smooths out the slower structures shed downstream. RAFT-CloudCav follows the ground truth most closely. It keeps the liquid--vapor interface sharp where the velocity gradient is steepest, resolves the thin low-velocity band inside the attached cavity, and recovers the shed structures near the right edge of the frame.

Averaged over the $183$ synthetic test pairs, which span all three frame intervals, the end-point error (EPE), the Euclidean distance between the estimated and true displacement vectors, is lowest for RAFT-CloudCav (table~\ref{tab:epe}). RAFT-CloudCav reaches $0.152$~px, about half the $0.300$~px of RAFT-Sintel. OpenPIV comes close at $0.185$~px. The liquid in these images is seeded with ideal particle images on a uniform background, which is the case cross-correlation is designed for, so its low error here is expected. Farneb\"ack returns $0.458$~px. All methods are scored on the same pixels, leaving out an $8$-pixel band along the frame edges where OpenPIV returns no vectors, and the RAFT models are run with $24$ recurrent iterations. The pre-trained RAFT checkpoints and the choice of Sintel as the starting point for fine-tuning are compared in Appendix~\ref{app:pretrained}.

\begin{table}[h]
    \centering
    \caption{Mean end-point error on the $183$ synthetic test pairs, with an $8$-pixel border left out.}
    \label{tab:epe}
    \begin{tabular}{lc}
        \hline
        Method & EPE (px) \\
        \hline
        Farneb\"ack & $0.458$ \\
        OpenPIV & $0.185$ \\
        RAFT-Sintel (pre-trained) & $0.300$ \\
        RAFT-CloudCav (fine-tuned) & $0.152$ \\
        \hline
    \end{tabular}
\end{table}

A synthetic benchmark of this kind also carries an inherent caveat that bears on how much weight it should take. The test pairs come from the same simulation as the fine-tuning data, although from later time steps, so RAFT-CloudCav is still evaluated close to its training distribution, and the same pairs were used to compare the starting checkpoints (Appendix~\ref{app:pretrained}). The remainder of this section therefore rests on experimental data, where no such advantage exists.

\subsection{Instantaneous Fields in Cloud Cavitation}
\label{sec:instantaneous}
\begin{figure}[p]
    \centering
    \includegraphics[width=0.99\linewidth,height=0.72\textheight,keepaspectratio]{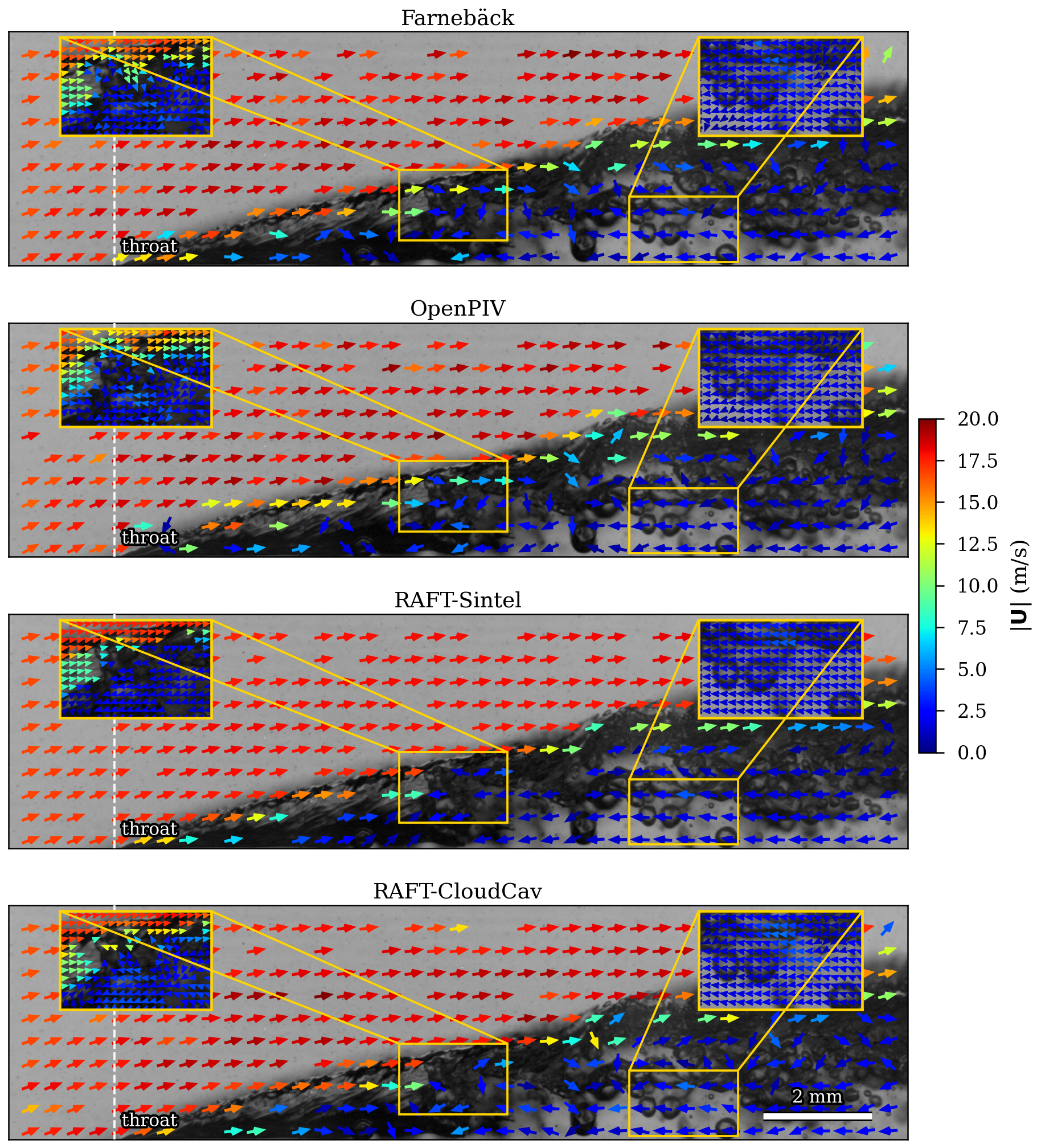}
    \caption{Instantaneous velocity fields at $48\text{ L/min}$ from a single frame pair, fine imaging configuration, overlaid on the source image. Panels run from classical to learned, in the order Farneb\"ack, OpenPIV, RAFT-Sintel and RAFT-CloudCav. Every panel uses the same frame, decimation, arrow scale, color limits, and the same two magnified regions, which show a recirculating cell at the cavity closure (left inset) and the re-entrant jet beneath the cavity (right inset). Arrows are drawn at unit length and colored by speed. A recirculating cell sits where the flow is nearly stagnant, so arrows scaled by magnitude would vanish exactly where the structure is. Direction is therefore carried by the arrow and magnitude by the color. Vectors below $0.4\text{ m/s}$ are omitted rather than assigned a meaningless direction. The dark region is the vapor cloud. A $2$~mm scale bar and the throat position are marked.}
    \label{fig:vectors}
\end{figure}

For a single frame pair at $48\text{ L/min}$, the four estimators return flow fields that agree in their main features and differ in their regularity (figure~\ref{fig:vectors}). All four panels use the same frame, vector spacing, arrow scale, and color limits, so they differ only by estimator.

Every method returns a dense field with a liquid free stream near $17\text{--}20\text{ m/s}$ above the cavity, a shear layer along the cavity interface, and slow flow of a few meters per second within the vapor cloud. The differences lie in how regular the field is rather than in what it contains. Farneb\"ack and OpenPIV scatter more in direction from one vector to the next, inside the cloud and along the interface, whereas both RAFT variants return fields that vary smoothly and follow the interface rather than crossing it.

The two magnified regions isolate the structures that define the shedding cycle. The right inset covers the near-wall layer beneath the cavity, where the re-entrant jet runs upstream. All four methods return reverse flow at every vector in this region, with a mean streamwise velocity of $-1.5$ to $-2.0\text{ m/s}$. They differ in how consistently the vectors are oriented. The circular standard deviation of the vector direction is $3^\circ$ for RAFT-Sintel, $13^\circ$ for RAFT-CloudCav, $15^\circ$ for OpenPIV and $17^\circ$ for Farneb\"ack. The jet is a thin liquid film beneath a vapor cavity and carries few tracer particles, but in this recording the bubbles and the interface above it supply texture that every estimator can follow. Where the image carries little texture, a learned estimator's output is shaped by its prior, so the regularity of the RAFT fields is consistent with the physics but is not on its own evidence that they are more accurate.

The left inset covers the cavity closure, the region in which Kelvin--Helmholtz roll-up entrains liquid into the detaching cloud \citep{podbevsek_kelvin_2021}. All four methods resolve the same structure. The free stream passes overhead at $15\text{--}17\text{ m/s}$, the streamwise velocity averaged across the inset falls through zero at $y \approx 0.9\text{--}1.0\text{ mm}$, and beneath it the fluid runs back upstream at $1.2\text{--}2.2\text{ m/s}$, with an upwash of $3\text{--}5\text{ m/s}$ where the flow turns. In the turning region the classical fields again scatter in direction, while the RAFT fields turn smoothly from the free stream into the reverse flow.

\subsection{Time-Averaged Profiles in Cavitating and Non-Cavitating Flow}
\label{sec:profiles}
\begin{figure}[h!]
    \centering
    \includegraphics[width=0.98\linewidth]{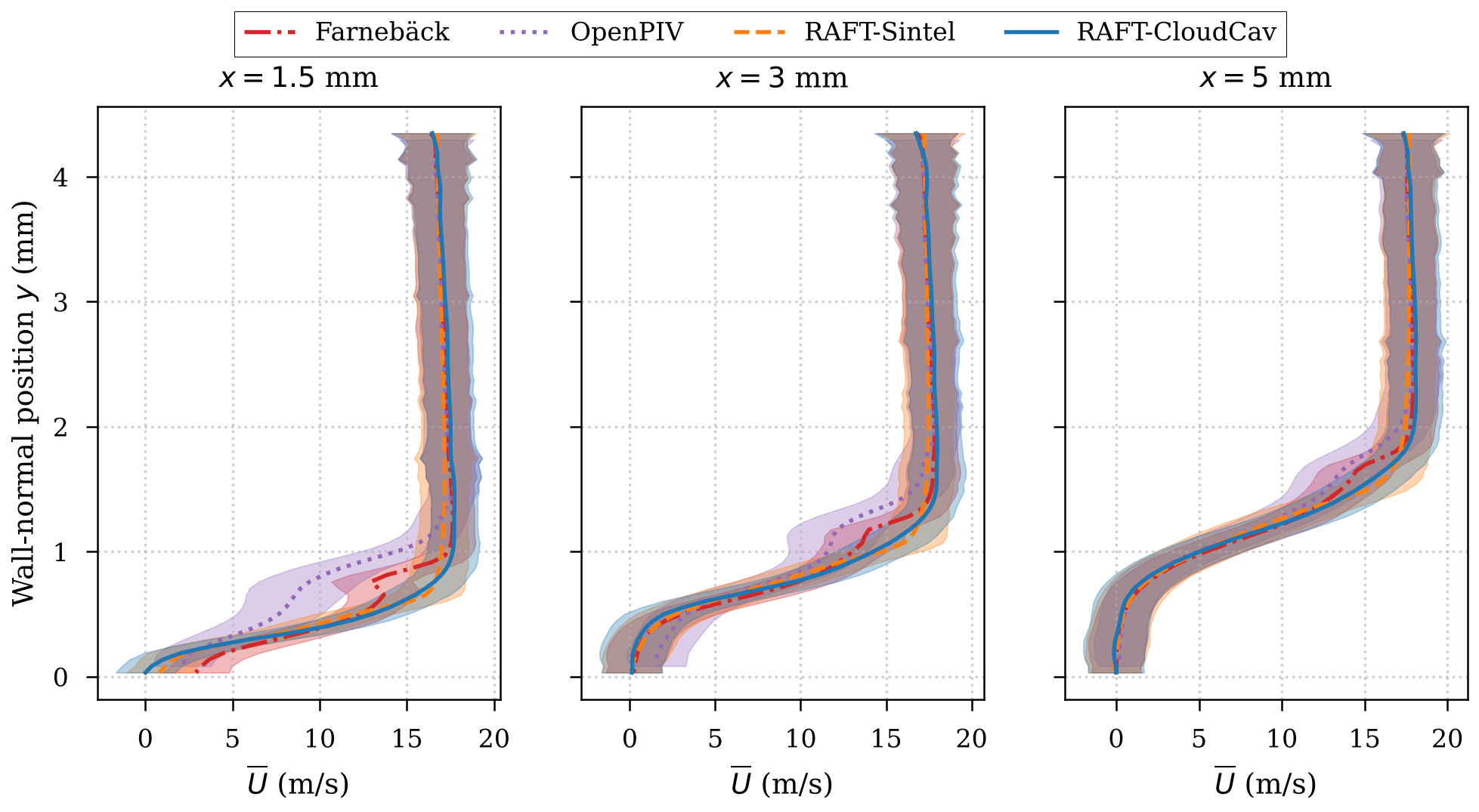}
    \caption{Mean streamwise velocity profiles at $48\text{ L/min}$ (cloud cavitation), fine configuration, at three stations downstream of the throat. Shaded bands are $\pm\sigma_{\bar{u}}$ as defined in Section~\ref{sec:uq}.}
    \label{fig:methods48}
\end{figure}

\begin{figure}[h!]
    \centering
    \includegraphics[width=0.98\linewidth]{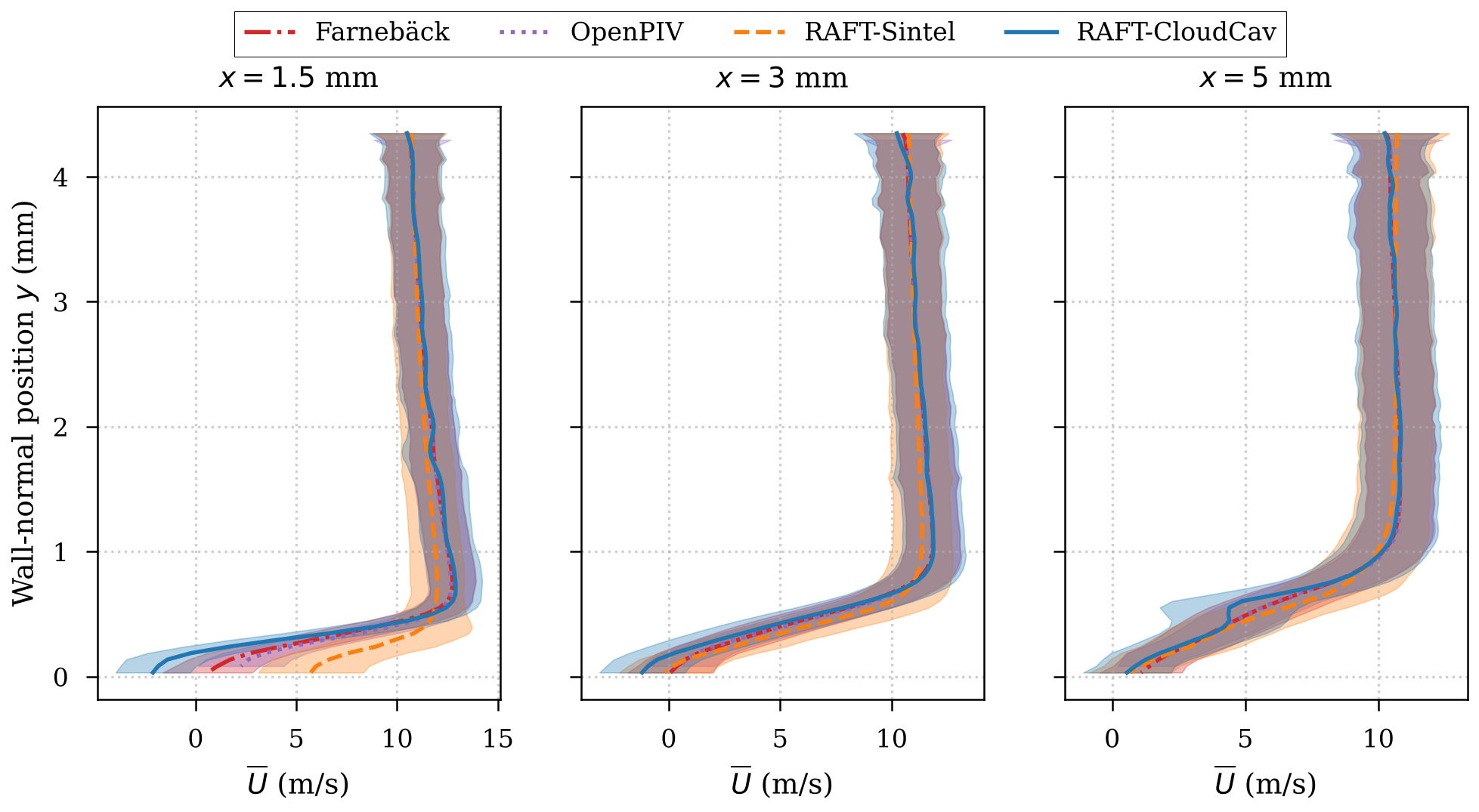}
    \caption{Mean streamwise velocity profiles at $30\text{ L/min}$ (non-cavitating control), fine configuration. Shaded bands are $\pm\sigma_{\bar{u}}$ as defined in Section~\ref{sec:uq}.}
    \label{fig:methods30}
\end{figure}

Mean streamwise velocity profiles at three downstream stations, $x = 1.5$, $3$ and $5$~mm, averaged over all $13{,}623$ frame pairs in the fine configuration, are compared at both operating points (figures~\ref{fig:methods48} and \ref{fig:methods30}). The $30\text{ L/min}$ case functions as a control, because there is no cavitation, the tracer particles are visible throughout, and classical methods operate inside their design envelope. Any difference there cannot be attributed to cavitation.

The control shows that the four methods agree where all of them operate within their design envelope. At $30\text{ L/min}$ and $x = 1.5$~mm they return profile-mean streamwise velocities of $10.66$, $11.00$, $10.80$ and $11.00\text{ m/s}$ for RAFT-CloudCav, RAFT-Sintel, Farneb\"ack and OpenPIV, within $3.2\%$ of one another. Most of that spread lies in the three vector rows nearest the wall, which fall within $12$ pixels of the frame edge (Section~\ref{sec:uq}). There the methods disagree by up to $7.8\text{ m/s}$, from $-2.1\text{ m/s}$ for RAFT-CloudCav, in a flow where the other three methods read positive values, to $+5.7\text{ m/s}$ for RAFT-Sintel. Values in those rows are treated as unreliable for every method.

At $48\text{ L/min}$ the methods agree in the liquid core and separate near the wall. At $x = 1.5$~mm and $y = 2\text{--}3$~mm, above the cavity, all four return $17.04\text{--}17.36\text{ m/s}$, within $1.9\%$ of one another. Over $y = 0.5\text{--}1$~mm, in the shear layer at the edge of the attached cavity, RAFT-CloudCav and RAFT-Sintel return $15.92$ and $15.74\text{ m/s}$, Farneb\"ack $14.14\text{ m/s}$ and OpenPIV $9.82\text{ m/s}$, $38\%$ below RAFT-CloudCav. This is where the cavity excludes tracer particles and the interface distorts the images of those that remain, the condition under which cross-correlation is expected to fail, and the OpenPIV deficit persists downstream to $x = 3$~mm (figure~\ref{fig:methods48}).

Taking the most negative point of the mean profile within $0.6\text{ mm}$ of the wall at $x = 5\text{ mm}$, the two RAFT models return a weak mean reverse flow, $-0.14\text{ m/s}$ for both RAFT-CloudCav and RAFT-Sintel, while Farneb\"ack returns $+0.01\text{ m/s}$ and OpenPIV $+0.19\text{ m/s}$. These values lie in the three rows nearest the wall, where RAFT-CloudCav reads $-2.1\text{ m/s}$ in the non-cavitating control, and well inside the conservative uncertainty of $1.5\text{--}1.9$~m/s. Even in the opposite limit, in which the random part of the residual averages down over $N_{\text{eff}} \approx 110\text{--}210$ independent samples to $0.13\text{--}0.22$~m/s, none of them is separated from zero. The time-averaged profiles therefore do not resolve the re-entrant jet. The jet is present for only part of each shedding cycle, and its mean signature is small against its instantaneous velocity of about $-2\text{ m/s}$ (Section~\ref{sec:instantaneous}), so the jet is better studied in the instantaneous fields than in the mean.

\subsection{Mass Conservation in the Liquid Phase}
\label{sec:divergence}
\begin{figure}[h!]
    \centering
    \includegraphics[width=0.98\linewidth]{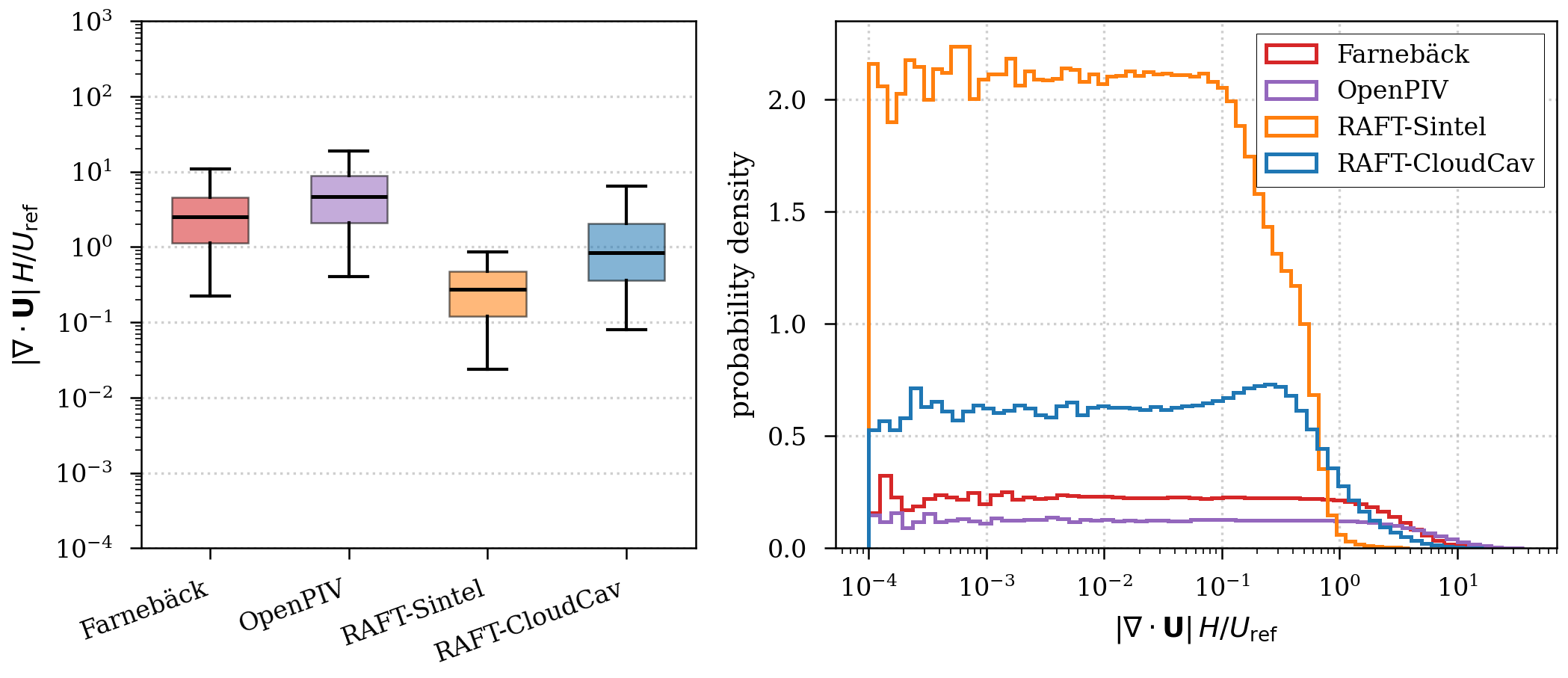}
    \caption{Normalized velocity divergence $|\nabla\cdot\mathbf{U}|H/U_{\text{ref}}$ over liquid-phase points at $48\text{ L/min}$, fine configuration, sampled over $200$ frames. The left panel shows the distribution by method and the right panel the probability density on logarithmic axes. Boxes span the interquartile range, with the median marked, and whiskers the 5th to 95th percentiles. Mass conservation requires this quantity to vanish.}
    \label{fig:div48}
\end{figure}

\begin{figure}[h!]
    \centering
    \includegraphics[width=0.98\linewidth]{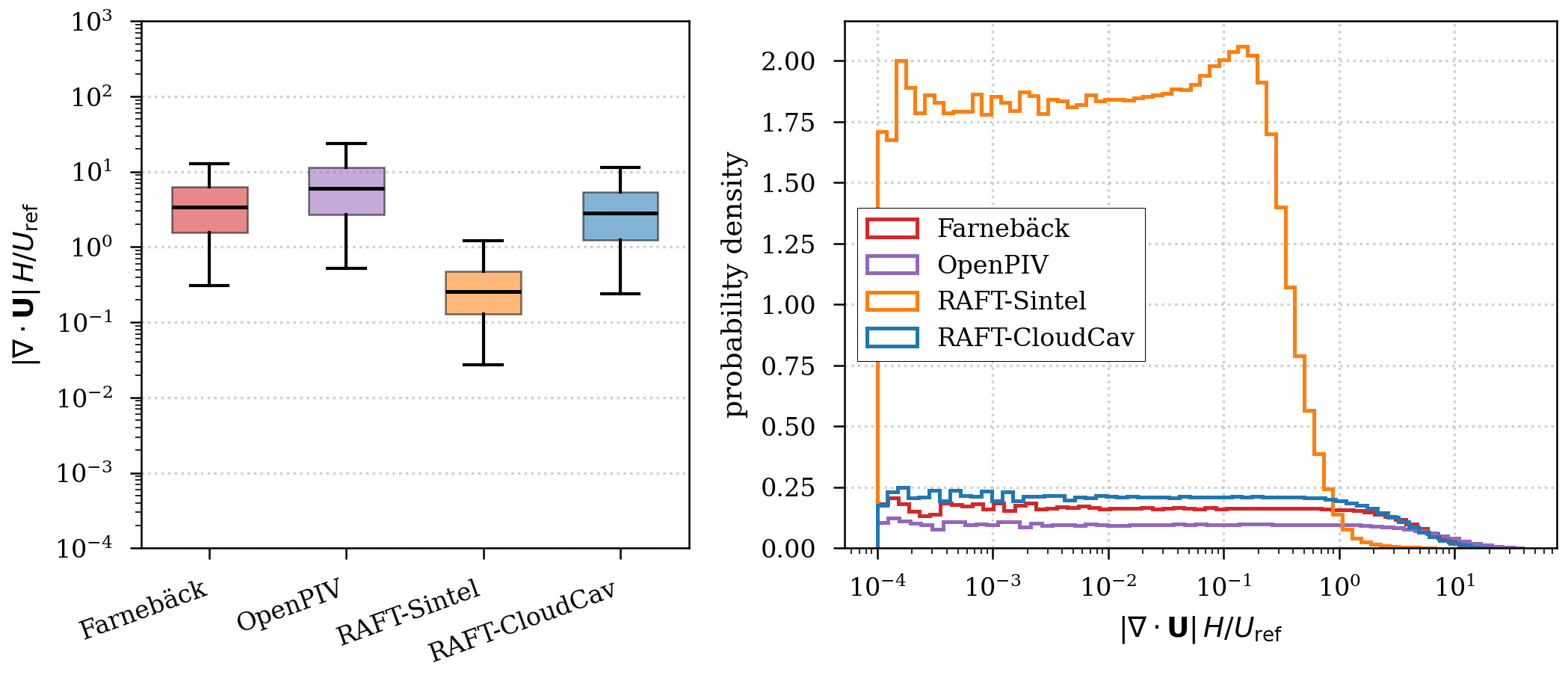}
    \caption{Normalized velocity divergence at $30\text{ L/min}$. With no vapor present the entire field is liquid, so no phase mask is required. Boxes and whiskers as in figure~\ref{fig:div48}.}
    \label{fig:div30}
\end{figure}

Both learned estimators satisfy continuity substantially better than the classical ones in the liquid phase, at both operating points (figures~\ref{fig:div48} and \ref{fig:div30}). At $30\text{ L/min}$ the absence of vapor makes the whole field liquid.

At $48\text{ L/min}$ the liquid mask retains $64\%$ of the field. The two learned estimators reach median normalized divergences of $0.84$ (RAFT-CloudCav) and $0.27$ (RAFT-Sintel), against $2.50$ for Farneb\"ack and $4.64$ for OpenPIV, and root-mean-square values of $3.25$ and $0.97$ against $6.27$ and $9.24$. The reference velocities agree to within $5\%$ across the four methods ($18.1\text{--}19.0$~m/s), so the normalization does not favor any of them. RAFT-Sintel is lowest on both measures. At $30\text{ L/min}$ the ordering is the same but the gap between the two learned models widens, with medians of $0.26$ for RAFT-Sintel, $2.81$ for RAFT-CloudCav, $3.43$ for Farneb\"ack and $5.99$ for OpenPIV. On this metric domain adaptation brings no gain. The difference does not come from the training displacement range, since the training pairs span the displacements of all three configurations (Section~\ref{sec:synthetic}). In the coarse configuration at $48\text{ L/min}$, where the displacements are smallest, the ordering is the same, with medians of $0.13$ for RAFT-Sintel, $0.30$ for RAFT-CloudCav, $0.79$ for Farneb\"ack and $0.88$ for OpenPIV. Nor does it come from the frame edges, since leaving out the $12$-pixel band along every edge changes the $30\text{ L/min}$ medians by less than $3\%$. RAFT-CloudCav simply returns a field with more small-scale variation than RAFT-Sintel, consistent with the larger direction scatter of its vectors in the jet (Section~\ref{sec:instantaneous}), and the caveat below applies to the comparison between the two.

One caveat applies to this metric in general. Divergence is a spatial derivative, so it amplifies high-frequency noise, and a method that returns a smoother field scores better almost automatically. That is partly the intended effect, since spurious vectors are precisely the failure being measured, but it means the metric alone cannot distinguish physical consistency from over-smoothing. It should therefore be read together with the instantaneous fields of Section~\ref{sec:instantaneous} and the near-wall profiles of Section~\ref{sec:profiles}, where over-smoothing would show as a weakened jet or a thickened shear layer.

\subsection{Integrated Flow Rate}
\label{sec:flowrate}
Integrated over the coarse configuration, the only one that observes almost the full channel height, all four methods fall $10\text{--}13\%$ below the meter and agree with one another to within $4\%$ (table~\ref{tab:flowrate}).

\begin{table}[h]
    \centering
    \caption{Integrated flow rate at the throat, and its variation across four streamwise stations ($x = 0$, $1.5$, $3$, $5\text{ mm}$ downstream of the throat). The station spread is the full range of the four station values normalized by their mean, $(Q_{\max} - Q_{\min})/\bar{Q}$, which mass conservation would drive to zero in a fully liquid flow. Rows run from classical to learned. Coarse configuration at $48\text{ L/min}$, with a metered value of $48.0 \pm 0.5\text{ L/min}$.}
    \label{tab:flowrate}
    \begin{tabular}{lccc}
        \toprule
        Method & $Q$ at throat [L/min] & Deviation & Station spread \\
        \midrule
        Farneb\"ack & $43.21 \pm 1.90$ & $-10.0\%$ & $4.4\%$ \\
        OpenPIV & $42.91 \pm 2.01$ & $-10.6\%$ & $4.2\%$ \\
        RAFT-Sintel & $41.56 \pm 2.85$ & $-13.4\%$ & $4.5\%$ \\
        RAFT-CloudCav & $41.99 \pm 2.76$ & $-12.5\%$ & $\mathbf{1.8\%}$ \\
        \bottomrule
    \end{tabular}
\end{table}

No method falls inside the meter's tolerance. The shortfall is shared by four estimators that work on different principles and agree with one another in the liquid core (Section~\ref{sec:profiles}), so it lies in the measurement rather than in any one estimator, and the data identify part of it. In the liquid core at $x = 1.5$~mm and $y = 2\text{--}3$~mm, the coarse configuration reads below the fine one by $2.8\%$ for OpenPIV, $4.5\%$ for Farneb\"ack, $7.2\%$ for RAFT-CloudCav and $9.7\%$ for RAFT-Sintel, whereas the fine and medium configurations agree within $0.5\%$ for RAFT-CloudCav. The coarse configuration is the one that spans the channel, so this deficit enters the flow rate directly and accounts for between a quarter and three quarters of it, depending on the method. Its cause is not established. An error in the calibration would scale every method equally, so the spread between methods points to an estimator effect at this resolution as well, larger for the two learned models, whose features are computed at $1/8$ resolution. The remainder of the shortfall is consistent with the projection through the channel depth (Section~\ref{sec:setup}), in which the depth of field weights the measurement toward the plane of focus at the front face of the channel, where the side-wall boundary layer slows the flow, and with the $3\%$ of the channel height at the throat that lies outside the field of view and is filled by the linear closure.

The uncertainties in table~\ref{tab:flowrate} inherit the conservative treatment of Section~\ref{sec:uq} and are propagated through the integral assuming the residual is coherent across the whole profile, which is the worst case, since a bias that changes sign across the shear layer would partly cancel. Even so, every method lies more than two standard deviations below the meter, consistent with a shortfall that does not come from image matching. Between methods the point estimates differ by at most $1.65\text{ L/min}$ and are not separated.

The station-to-station spread does not depend on the uncertainty model at all. RAFT-CloudCav varies by $1.8\%$ across four stations spanning $5\text{ mm}$ of the divergent section, while the other three methods vary by $4.2\text{--}4.5\%$. Each of those three reads lowest at $x = 5$~mm, where the cavity is thickest, by $2.5\text{--}4.3\%$ below its value at $x = 1.5$~mm, against $1.7\%$ for RAFT-CloudCav. As noted in Section~\ref{sec:metrics}, the downstream stations cut through the cavity, where the integral is not a strict mass balance, so this is a measure of how consistently each method integrates the same flow from station to station rather than a direct test of mass conservation. It is the one integral measure on which domain adaptation gives a clear gain.

\subsection{Spatial Resolution}
\label{sec:spatial}
\begin{figure}[h!]
    \centering
    \includegraphics[width=0.98\linewidth]{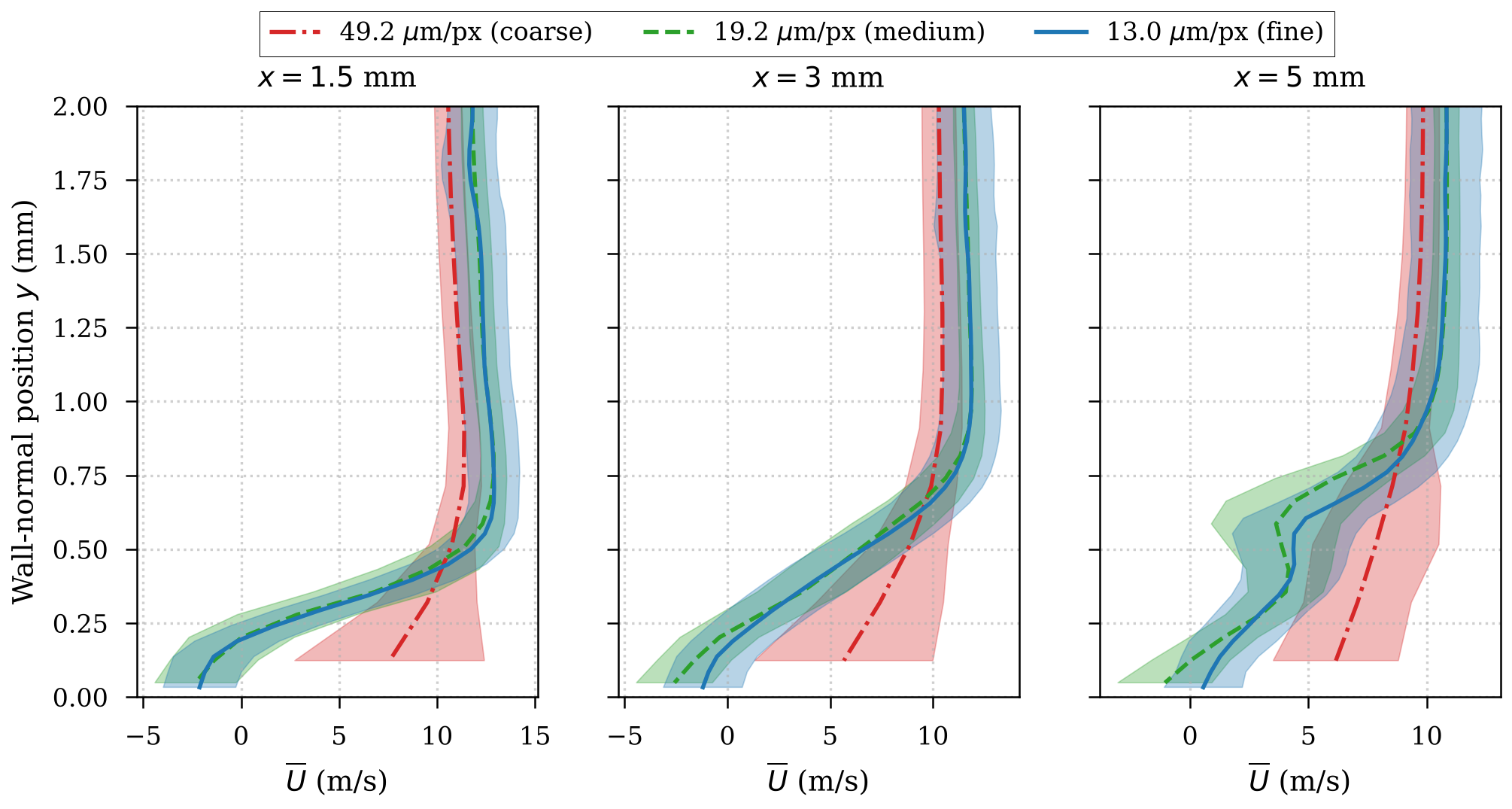}
    \caption{Mean streamwise velocity from RAFT-CloudCav across three imaging configurations at $30\text{ L/min}$, resolved onto the bottom-wall boundary layer at three downstream stations. Shaded bands are $\pm\sigma_{\bar{u}}$. The coarse configuration begins at $y = 0.12\text{ mm}$. The three points nearest the wall in each configuration lie within $12$ pixels of the frame edge, below $y = 0.16$, $0.23$ and $0.59$~mm for the fine, medium and coarse configurations.}
    \label{fig:spatial}
\end{figure}

Across the three imaging configurations at $30\text{ L/min}$, the two finer ones return the same near-wall profile while the coarse one departs from both (figure~\ref{fig:spatial}, resolved onto the wall-normal region containing the boundary layer). The non-cavitating condition is used deliberately. With no vapor present, the near-wall region is a clean boundary layer on the bottom wall rather than a cavity interface, so a difference between configurations can be attributed to sampling rather than to two-phase physics.

The decisive comparison is not the velocity but the wall-normal gradient, because it does not depend on the uncertainty model. At $x = 1.5$~mm the boundary layer thickness, taken where the velocity first reaches $99\%$ of the profile maximum, is $\delta_{99} \approx 0.66\text{ mm}$, and the three configurations sample it with grid spacings of $0.052$, $0.077$ and $0.197\text{ mm}$, that is, with $12$, $8$ and $3$ points inside the layer. Taking the largest central difference of the measured profile, without imposing the no-slip condition, the fine and medium configurations recover peak gradients of $46.5$ and $43.7\ (\text{m/s})/\text{mm}$, at $y = 0.29$ and $0.36$~mm, leaving out stencils that use a point within $12$ pixels of the frame edge. The coarse configuration recovers only $8.1\ (\text{m/s})/\text{mm}$, an under-prediction by a factor of nearly six, from a stencil that spans $0.12$ to $0.52$~mm. Imposing no-slip would change this picture, since the wall-to-first-point gradient of the coarse profile is $62\ (\text{m/s})/\text{mm}$. The measured gradients describe what each configuration resolves above its first grid point.

The mechanism is visible in the raw numbers rather than inferred. The first grid point of the coarse configuration sits at $y = 0.12\text{ mm}$ and already reports $7.58\text{ m/s}$, which is $67\%$ of that configuration's own maximum of $11.39\text{ m/s}$. The fine and medium profiles reach that velocity only at $y \approx 0.37$~mm, so the coarse profile is displaced toward the wall by about a quarter of a millimeter, five pixels at $49.2~\mu$m/pixel. The coarse configuration does not so much under-resolve the boundary layer as step over it. Grid spacing alone does not explain this. RAFT also operates on a feature map at $1/8$ resolution, $0.39$~mm at $49~\mu$m/pixel, more than half the layer thickness, and a one-pixel error in the wall position is $0.05$~mm. The coarse result is therefore a combination of sampling, the estimator's effective resolution and wall registration, which these data do not separate. Nearest the wall the finer configurations are no more reliable. Their first three points, within the edge band, read $-2.3$ to $0.0\text{ m/s}$, where the other estimators read positive values (Section~\ref{sec:profiles}), so the profile below about $0.2$~mm is not measured reliably in any configuration. Away from the wall the coarse configuration also reads $9\%$ below the other two, at $y = 2\text{--}3$~mm, the same deficit that enters the flow rate of Section~\ref{sec:flowrate}.

Viewed through the conservative uncertainty, the comparison is not resolved. Over $y = 0.25\text{--}1.5\text{ mm}$ the medium and coarse profiles differ by up to $7.0\text{ m/s}$ but by at most $1.6\sigma$, while the fine and medium profiles differ by at most $0.57\text{ m/s}$ ($0.3\sigma$). The boundary layer is therefore captured better at each step from coarse to fine, with $3$, $8$ and $12$ points across it and peak gradients of $8.1$, $43.7$ and $46.5\ (\text{m/s})/\text{mm}$. The coarse configuration does not resolve it, and the two finer ones agree with each other within $6\%$ in peak gradient, so the last refinement adds points inside the layer more than it changes the gradient.

\subsubsection{Uncertainty near the wall and in the free stream.}
Near the wall the reported uncertainty also falls as the imaging is refined. Taking the largest $\sigma_{\bar{u}}$ inside the boundary layer ($y \le \delta_{99}$) at $x = 1.5$~mm, it is $2.30$, $3.13$ and $4.84\text{ m/s}$ for the fine, medium and coarse configurations. The coarse maximum sits at its first grid point, $y = 0.12$~mm, where a single $4$-pixel block spans $0.20$~mm of the steepest part of the layer. All three coarse points inside the layer also lie in the $12$-pixel edge band (Section~\ref{sec:uq}), although in the finer configurations the edge rows do not carry the largest values. A finer grid therefore returns both more points inside the boundary layer and a smaller uncertainty at each of them.

Away from the wall the ordering is different. The median of $\sigma_{\bar{u}}$ across the whole profile is $1.35$, $0.66$ and $0.64\text{ m/s}$, highest in the fine configuration, and in pixel units the residual falls as the imaging coarsens, at $0.80$, $0.26$ and $0.10\text{ px}$. Most of this is attributed to the test-section vibration of Section~\ref{sec:setup}. A vibration of fixed physical amplitude spans the most pixels at $13$~$\mu$m/pixel, and the part of it that is not a uniform translation raises the residual. The displacement per frame contributes as well, since peak displacements at this condition are $7.9$, $5.3$ and $1.9\text{ px/frame}$ and a larger displacement returns a larger residual, and so may the particle images, which measure $0.8$~pixels in the fine and coarse configurations against $2.1$ in the medium one (Section~\ref{sec:setup}). Converting a pixel residual to a velocity then multiplies it by the calibration, $1.69\text{ m/s}$ per px/frame at $13$~$\mu$m/pixel but $6.39\text{ m/s}$ at $49.2$~$\mu$m/pixel (both at $130{,}000$~frames/s), which brings the medium and coarse configurations level in the free stream.

Finer imaging is therefore the better choice for the near-wall region, and the resolution can be chosen from the fidelity that an application needs there. What finer imaging demands in return is mechanical stability. In cloud cavitation the collapse of shed clouds excites the test section, so the vibration has to be mitigated before the magnification is raised, or the gain near the wall comes with a larger uncertainty in the free stream.

\subsection{Seeded and Unseeded Operation}
\label{sec:seeding}
\begin{figure}[h!]
    \centering
    \includegraphics[width=0.98\linewidth]{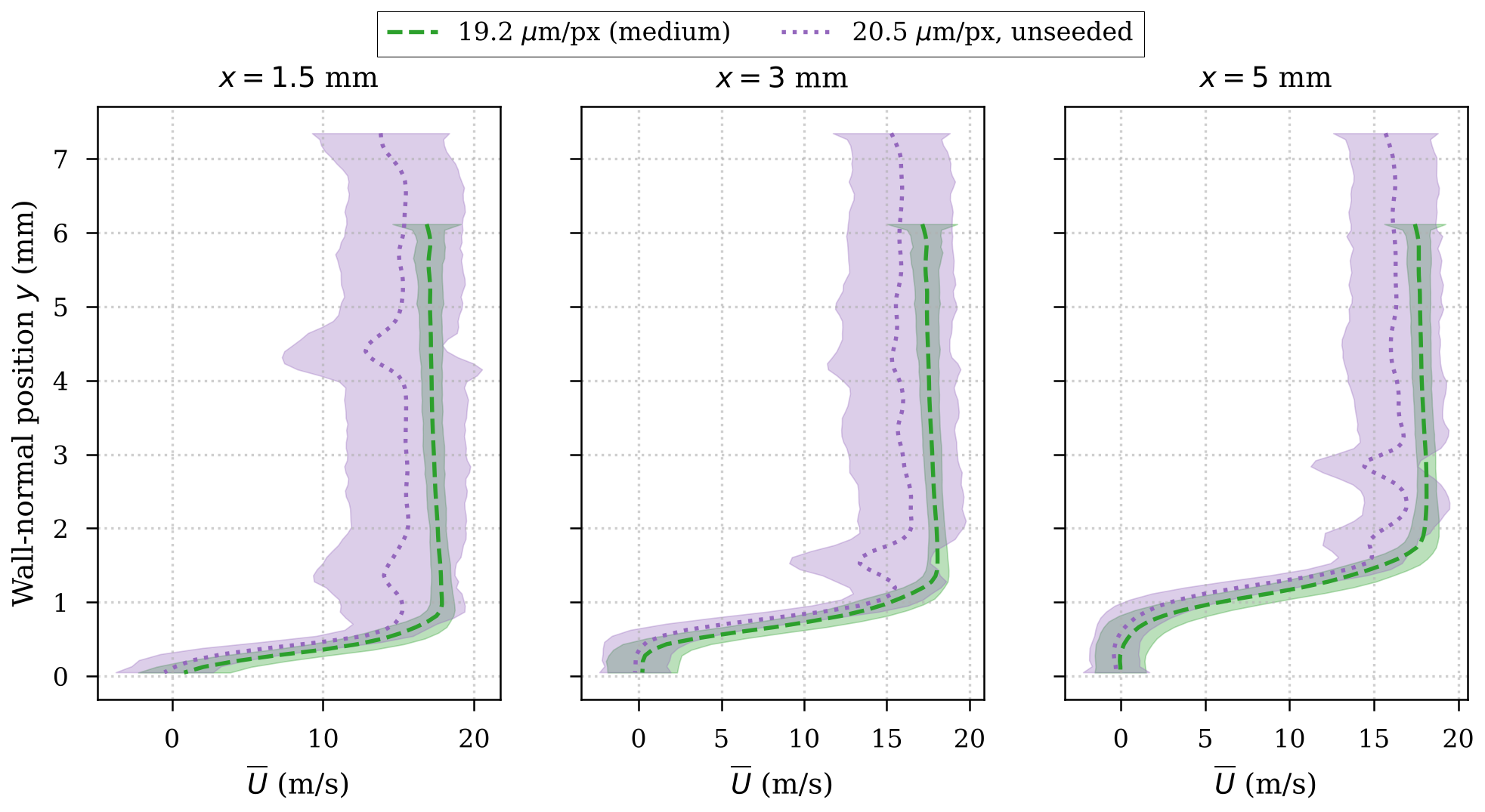}
    \caption{RAFT-CloudCav on seeded and unseeded recordings at $48\text{ L/min}$ and comparable magnification. Shaded bands are $\pm\sigma_{\bar{u}}$.}
    \label{fig:seeding}
\end{figure}

RAFT-CloudCav recovers the shape of the mean profile with and without tracer particles, at comparable magnification (figure~\ref{fig:seeding}). The two recordings are not at identical calibration. The seeded case is the medium configuration at $19.2$~$\mu$m/pixel, seeded with $40~\mu$m particles, and the unseeded case was acquired at $20.5$~$\mu$m/pixel, a $7\%$ difference in pixel size. The comparison is therefore between two closely matched configurations rather than one, and the residual calibration difference is not separated from the seeding effect. The profiles agree in shape but not fully in level. Averaging over the liquid core at $x = 1.5$~mm, taken as $y > 2.0\text{ mm}$, which lies above the cavity at this station, the mean streamwise velocity is $17.21\text{ m/s}$ seeded against $15.05\text{ m/s}$ unseeded, a difference of $12.6\%$ that lies inside the unseeded uncertainty, and the shear layer is recovered in both. The uncertainty averaged over the same region is about five times larger without seeding ($4.15$ against $0.76\text{ m/s}$), which is the substantive cost of relying on natural void texture rather than a controlled tracer distribution. The mean field survives, but the confidence in any single frame does not.

The physical origin of the difference is straightforward. In the liquid region away from the cavity there is little natural texture to track, and an estimator that relies on persistent image structure has correspondingly less to work with there. That raises the question of what the unseeded estimate in that region is built on. Some of it is natural micro-bubbles and impurities, and some is likely carried in from the textured vapor region by the estimator's spatial coherence. The two sources are not separated here. The two recordings were also made separately, and the unseeded one did not vibrate, so differences in operating condition between the runs are not separated from the effect of seeding either.

This is a practical result rather than a marginal one. Seeding a cavitation tunnel is invasive. Tracer particles can act as heterogeneous nucleation sites, because gas trapped in crevices on their surfaces lowers the tension needed for inception \citep{atchley_crevice_1989}, and so they can alter the cavitation behavior being measured \citep{marschall_cavitation_2003, borkent_acceleration_2008}. Smaller, smoother particles offer fewer surface irregularities to trap nuclei, but they do not remove the effect. Introducing and removing seeding also adds labor. That a domain-adapted optical flow estimator recovers the mean field to within $13\%$ from natural texture alone removes a constraint that PIV cannot relax, since cross-correlation has nothing to correlate without particles.

\section{Operational Guidelines}
\label{sec:guidelines}
The sensitivity studies of Sections~\ref{sec:profiles} to \ref{sec:seeding} bound an operating envelope for applying a fine-tuned optical flow model to a comparable two-phase flow. The recommendations are collected in table~\ref{tab:guidelines} and the reasoning behind each is set out below. They are drawn from one facility and one geometry, so they should be read as a starting point for an experiment of this kind rather than as validated limits.

The binding constraint on near-wall accuracy is sampling, not the estimator. A boundary layer is recovered only if enough grid points fall inside it, and the vector grid is set by the imaging calibration and the block size together. At $30$~L/min the layer at $x = 1.5$~mm is $\delta_{99} \approx 0.66$~mm thick, and the three configurations place $12$, $8$, and $3$ points inside it. The first two agree on the peak gradient to within $6\%$ ($46.5$ against $43.7\ (\text{m/s})/\text{mm}$) while the third under-predicts it nearly sixfold, because its first grid point already sits at $67\%$ of its maximum velocity (Section~\ref{sec:spatial}). Finer imaging captured the layer better at every step, with more points inside it and a smaller near-wall uncertainty, so the resolution should be chosen from the near-wall fidelity that the application needs. A vector spacing of about $\delta_{99}/8$, about $20$~$\mu$m/pixel here for the $4$-pixel block, was enough to recover the peak gradient, and it scales with the boundary layer being measured. Finer imaging also raised the uncertainty in the free stream, which is attributed mainly to test-section vibration (Section~\ref{sec:spatial}), so it pays off fully only when the test section is mechanically stable. Separately, the three vector rows within $12$ pixels of the edge of the processed frame, which here coincides with the wall, were unreliable for every estimator (Section~\ref{sec:profiles}). That suggests cropping so that the wall lies inside the processed frame rather than on its edge, which was not tested here.

Integral quantities impose a separate and independent constraint, because they depend on the extent of the field of view rather than on its resolution. A flow rate obtained from a profile that covers less than the channel height is an extrapolation over the unmeasured part, and the near-wall region that is missing is exactly where the profile departs from uniform. The fine configuration sees $44\%$ of the channel height at the throat and the coarse one $99\%$ (table~\ref{tab:configs}), which is why only the coarse configuration is used for the flow rate of Section~\ref{sec:flowrate} despite resolving the boundary layer worst. The requirement also tightens downstream. The channel opens as $H(x) = H + x\tan 8^\circ$, so a field of view that spans the throat covers only $93\%$ of the channel five millimeters further along. An experiment that needs both a resolved near-wall profile and a measured flow rate needs two recordings, not one, and the two should be compared in the core, where here the wide-field recording read $3\text{--}10\%$ below the finer one (Section~\ref{sec:flowrate}).

Displacement per frame and mechanical stability act on the same quantity, the pixel residual, and both worsen as magnification increases. Peak displacements here were $7.9$, $5.3$, and $1.9$~px/frame across the three configurations, and the pixel residual tracked them at $0.80$, $0.26$, and $0.10$~px. Magnification and frame rate must therefore be raised together, keeping the displacement below roughly $10$~px/frame. At $48$~L/min the fine configuration reaches $10.9$~px/frame, at the edge of this guideline and of the training set, which was rendered at three frame intervals to span the displacements of all three configurations (Section~\ref{sec:synthetic}). The training displacement range should be matched to the experiment in this way, or the experiment to it. Vibration compounds this. A rigid displacement of the test section registers as a whole-field translation of fixed physical amplitude, so its size in pixels scales inversely with the calibration. Its uniform part is tracked as real motion and biases the velocity, and the rest raises the residual, which is the likely main cause of the larger free-stream uncertainty of the fine configuration. Cloud cavitation makes this harder, because the collapse of shed clouds excites the test section. The practical order of operations is to isolate the test section first and increase magnification afterwards, and to measure any remaining vibration from a static feature in the image.

Illumination and seeding are the two choices that determine how much trackable texture reaches the sensor. Oblique illumination at roughly $45^\circ$ is required rather than preferred, because backlighting reduces the vapor to a silhouette and removes the interior texture that the estimator tracks, which is the difference visible in figure~\ref{fig:sample-frames}. Seeding, by contrast, turns out to be optional for mean fields. RAFT-CloudCav recovered the core velocity to within $13\%$ without tracer particles, at a per-frame uncertainty about five times larger ($4.15$ against $0.76$~m/s), because the liquid away from the cavity offers little natural texture (Section~\ref{sec:seeding}). Mean profiles and integral quantities survive that trade, but instantaneous fields and any turbulence statistic do not, so seeding remains necessary when single-frame confidence is what is wanted.

Finally, domain adaptation changed how consistently the flow was recovered rather than how much of it. Fine-tuning on rendered pairs from a simulation of the same geometry halved the end-point error on synthetic pairs and gave the most consistent flow rate between stations, $1.8\%$ against $4.5\%$ for the generalist RAFT-Sintel weights. It did not reduce the common flow-rate shortfall, where the two RAFT models differ by less than $1\%$, and the generalist weights gave the lower liquid-phase divergence in every configuration tested, although the training displacements spanned all of them. The training data should still span the displacements and particle-image sizes of the experiment, and here they matched the displacements but not the particle-image sizes. The model here was trained and tested on a single Venturi, so it has not been shown to transfer. A new geometry, a substantially different cavitation regime, or a different imaging arrangement should be assumed to need retraining, and the ground-truth-free checks of Sections~\ref{sec:profiles} to \ref{sec:flowrate}, which need no reference velocity field, are the means of testing whether it does.

\begin{table}[htbp]
    \centering
    \caption{Recommended operating parameters for deep optical flow velocimetry in cavitating flows.}
    \label{tab:guidelines}
    \begin{tabular}{p{0.22\linewidth}p{0.24\linewidth}p{0.44\linewidth}}
        \toprule
        Parameter & Recommendation & Basis \\
        \midrule
        Spatial resolution & Vector spacing $\le \delta_{99}/8$ ($\approx 20$~$\mu$m/pixel here), finer as the near-wall fidelity requires & Each step from coarse to fine added points inside the boundary layer and lowered the near-wall uncertainty. $49\ \mu\text{m/px}$ gives $3$ points across $\delta_{99}$ and under-predicts the peak gradient nearly sixfold. Vectors within $12$ pixels of the frame edge were unreliable. \\
        Field of view & Must span the channel for integral quantities & The fine configuration observes $44\%$ of the channel height at the throat and the coarse one $99\%$ (table~\ref{tab:configs}). Only the latter supports a flow rate that is measured rather than extrapolated, and even it covers $93\%$ of $H(x)$ at the most downstream station. \\
        Displacement per frame & Keep below roughly $10\text{ px/frame}$ and within the training range & Pixel-level precision degrades as displacement grows, so magnification and frame rate must be raised together. \\
        Mechanical isolation & Isolate the test section before magnifying, and measure residual vibration from a static feature & Its uniform part biases the velocity without appearing in the uncertainty, and the rest raises the residual. Its pixel amplitude is largest in fine imaging, and cloud collapse excites it. \\
        Illumination & Oblique ($\approx 45^\circ$), not backlit & Backlighting reduces vapor to a silhouette and removes the texture the estimator tracks. \\
        Seeding & Optional for mean fields & Unseeded operation recovers the mean field to within $13\%$, at a per-frame uncertainty about five times larger, because the liquid region away from the cavity offers little natural texture to track. \\
        Model & Fine-tuned on data that span the experiment's displacements & Fine-tuning halved the synthetic end-point error and gave the most consistent flow rate between stations, but the generalist model gave the lower divergence. \\
        \bottomrule
    \end{tabular}
\end{table}

\section{Limitations}
\label{sec:limitations}
Several limitations bound the claims above and indicate where the method should be strengthened.

The uncertainty estimate has not been validated against ground truth. Its construction is physically motivated and its propagation is standard, but no coverage test has been performed, so the fraction of points at which the true error falls within $\pm\sigma$ is unmeasured. The synthetic dataset would support such a test directly, and until it is performed $\sigma$ should be read as an indicator rather than a calibrated interval.

More specifically, the signed residual shows that almost all of the residual fluctuates from frame to frame, and the variance argument of Section~\ref{sec:uq} shows that it overstates the velocity error, but neither shows by how much. The conservative limit is therefore adopted throughout, which is why the reported bands are wide enough that several comparisons establish an ordering without resolving a gap. At $x = 1.5$~mm and $30$~L/min the averaged limit is smaller by a factor of $23$, $13$ and $1.7$ in the fine, medium and coarse configurations, so the choice is consequential. The same excess of residual variance over measured variance means that the noise-inflation correction to the Reynolds stresses, which subtracts the random residual variance, would return negative stresses at most points, so it is withheld. A coverage test would settle both.

The uncertainty budget covers image matching only (table~\ref{tab:budget}). Calibration uncertainty, timing, out-of-plane motion through the $4.76\text{ mm}$ channel depth, and any bias from the estimator's learned smoothness prior that leaves no signature in the image residual, for example a gradient that is smoothed but still produces a well-matched warped image, are all excluded. Of these, out-of-plane motion is likely the largest for a flow with strong three-dimensional structure, and it is not addressed by a planar measurement at all.

The measurement is a projection rather than a plane. The test section is illuminated in volume, with no light sheet, so every reported vector is an average of the in-plane motion of all trackable texture across the $4.76\text{ mm}$ channel depth, weighted by the depth of field of the lens, which was not quantified because the aperture setting was not recorded. Two consequences follow. Spanwise variation within the cavity is averaged rather than resolved, which matters most near the closure region where the cavity is least two-dimensional. And within the vapor the estimator tracks image texture and interface motion, which need not equal the local liquid velocity, since a bubbly mixture and the liquid carrying it can move at different speeds, and interfacial features can propagate without material transport. Quantities reported inside the vapor should therefore be read as the motion of the visible structure, not as a liquid velocity \citep{khlifa_fast_2017}. The divergence metric is evaluated in the liquid phase for this reason. The flow-rate integral is not, and Section~\ref{sec:metrics} sets out what that implies for the station spread. The depth weighting is also a candidate for part of the flow-rate shortfall of Section~\ref{sec:flowrate}, since it favors the plane of focus at the front face of the channel, close to the side wall.

Two properties of the working fluid that bear on cavitation inception were not measured, as noted in Section~\ref{sec:setup}. Without the free-stream nuclei size distribution the absolute inception behavior cannot be compared against another facility even at the same cavitation number, and the effect of the tracer particles as nucleation sites (Section~\ref{sec:seeding}) cannot be quantified. The dissolved gas content was likewise not recorded.

The tracer particles follow the bulk flow but not every scale of it. For the $10~\mu$m particles of the fine configuration the Stokes number of Section~\ref{sec:setup} is $0.013$ against the bulk timescale but rises to $\approx 0.36$ against the peak near-wall shear, so the tracers lag somewhat in the steepest part of the near-wall profile. The $40~\mu$m particles of the medium and coarse configurations, needed to give resolvable particle images over the larger fields of view, reach $\mathrm{Stk} \approx 0.21$ against the bulk timescale and $\approx 5.8$ against the same shear, so they do not follow the steepest near-wall shear at all. The agreement between the $13$ and $19$~$\mu$m/pixel profiles is therefore between two configurations that differ in tracer size as well as in calibration, and the two effects are not separated. That is the same region on which the spatial-resolution result of Section~\ref{sec:spatial} rests, and it is one reason the unseeded result matters, since it recovers the shape of the mean field without relying on particles that are imperfect followers exactly where the gradient is largest.

There is no independent local velocity reference. The re-entrant jet is resolved by every method in the instantaneous field, but its time-averaged signature is smaller than the uncertainty and lies in the three vector rows nearest the wall, which are unreliable for every estimator because the wall coincides with the edge of the processed frame. The near-wall profile below about $0.2$~mm is therefore not measured, and the $10\text{--}13\%$ flow-rate shortfall common to all four methods is only partly explained. Laser Doppler velocimetry or X-ray velocimetry in the same facility would test both directly, and neither was available.

The model was trained and tested on a single Venturi geometry at one cavitation regime, so nothing here establishes that it transfers. Within that geometry its training displacements were matched to the experiment by rendering each plane at a different frame interval, but its particle images were not matched in size. The ground-truth-free checks used throughout are, however, exactly the tools with which transfer could be tested in a new facility without a reference field.

The divergence metric is subject to the smoothing confound noted above, and to the non-uniform depth weighting of the projected measurement. It compares methods on a common basis rather than measuring an absolute physical residual.

Finally, the synthetic test block comes from the same simulation as the training data, although from later time steps, so its end-point errors still favor RAFT-CloudCav and are not a measure of accuracy on real recordings. A held-out operating point or geometry would give a stricter estimate. The experimental results do not share this weakness, which is why the quantitative claims here rest on them.

\section{Conclusion}
This work presented a domain-adapted deep optical flow framework for velocimetry in unsteady cloud cavitation, and evaluated it against a pre-trained generalist model, a classical dense optical flow method, and cross-correlation PIV on high-speed Venturi recordings.

Validation was built on measurements requiring no reference velocity field. The four methods agreed within $2\%$ in the liquid core at $48\text{ L/min}$ and integrated to $41.6\text{--}43.2\text{ L/min}$ against a metered $48.0 \pm 0.5\text{ L/min}$. That common shortfall of $10\text{--}13\%$ lies in the measurement rather than in the estimators, and part of it was traced to the coarse configuration that spans the channel, which read $3\text{--}10\%$ below the finer ones in the core. RAFT-CloudCav gave the most consistent flow rate between streamwise stations, varying by $1.8\%$ against $4.2\text{--}4.5\%$ for the other methods. The learned estimators satisfied liquid-phase continuity three to seventeen times better than the classical ones, with the generalist RAFT-Sintel lowest, and OpenPIV under-read the cavitating near-wall shear layer by up to $38\%$. Every method resolved the re-entrant jet in the instantaneous field, the learned ones with less directional scatter, but its mean signature was not resolved in the time-averaged profiles. A non-cavitating control at $30\text{ L/min}$ confirmed that the four methods agree where all of them operate within their design envelope.

Sensitivity studies established the operating envelope collected in Section~\ref{sec:guidelines}. Imaging resolution limits the recoverable near-wall gradient through simple sampling. At $49$~$\mu$m/pixel only three grid points fall inside a $0.66\text{ mm}$ boundary layer and the peak gradient is under-predicted nearly sixfold, while $13$ and $19$~$\mu$m/pixel agree with each other within $6\%$. Finer imaging also lowered the near-wall uncertainty, so the resolution can be chosen from the near-wall fidelity required, provided the test section is isolated from the vibration that cloud collapse excites. Field of view constrains integral quantities independently of local accuracy, so a resolved near-wall profile and a measured flow rate can require separate recordings. Unseeded operation recovers the mean field to within $13\%$ at about five times the per-frame uncertainty, removing a constraint that cross-correlation cannot relax. The signed image residual showed the conservative uncertainty to be an upper bound, and calibrating the residual against ground truth is the main step needed to tighten it.

The measurement pipeline is released as an open-source package.

\section*{Statements and Declarations}

\paragraph{Funding} This research received no external funding.

\paragraph{Competing interests} The authors declare no competing interests.

\paragraph{Ethics approval} Not applicable.

\paragraph{Author contributions} Naga Nitish Chamala contributed the conceptualization, methodology, software, formal analysis, investigation, data curation and visualization, wrote the original draft, and reviewed and edited the manuscript. Sanjay Vasanth Kethanur Balasubramaniam contributed to the investigation through the experimental setup, prepared the illustrations of the experimental setup, and reviewed and edited the manuscript. Olivier Coutier-Delgosha provided the resources, supervised the work, and reviewed and edited the manuscript. All authors read and approved the final manuscript.

\paragraph{Data availability} The data that support the findings of this study are available from the corresponding author upon reasonable request.

\paragraph{Code availability} The velocimetry pipeline, including the uncertainty estimation and statistical reduction described in Section~\ref{sec:uq}, is openly available at \url{https://github.com/naganitishchamala/velocimetry_for_multiphase}.

\appendix

\section{Calibration Through the Channel Depth}
\label{app:calibration}
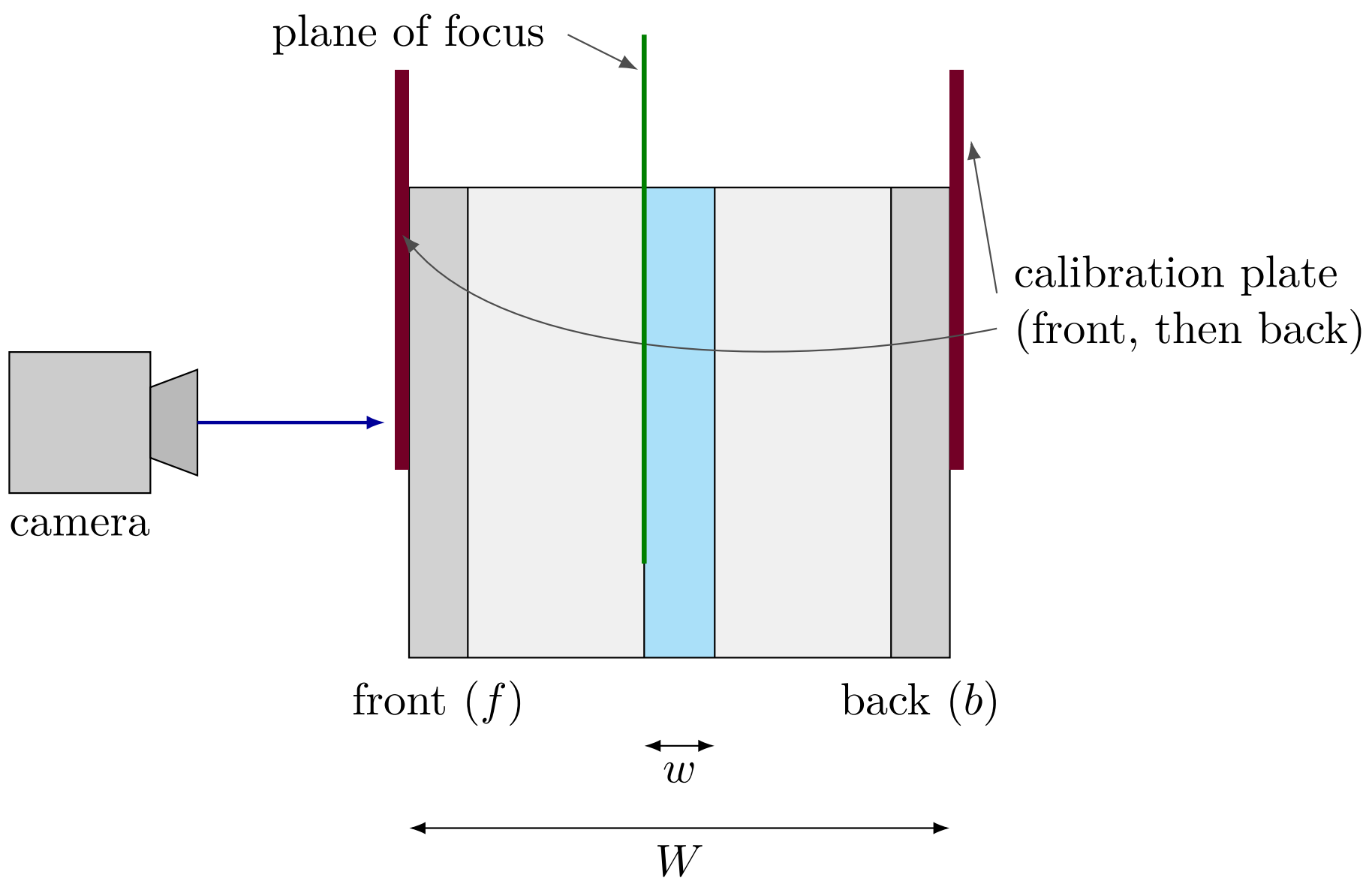
\begin{figure}[h!]
    \centering
    \resizebox{0.75\linewidth}{!}{\input{tikz/calibration}}
    \caption{Calibration through the depth of the test section, viewed from above. The dot-pattern plate is imaged against the front outer wall ($f$) and then against the back outer wall ($b$), and the scale factor at the plane of focus, the front face of the channel, is interpolated from the two. $w$ is the channel width and $W$ the distance between the outer walls.}
    \label{fig:calibration}
\end{figure}

The camera images the flow through an acrylic block, so the scale factor at the plane of focus cannot be measured by placing a target there. Instead, a LaVision calibration plate with dots at measured spacing was imaged twice, first against the front outer wall of the test section, giving the scale factor $C_f$, and then against the back outer wall, giving $C_b$ (figure~\ref{fig:calibration}). The plane of focus is the front face of the channel, and its scale factor is interpolated from the two measurements as
\begin{equation}
    C_p = \frac{2\,C_f C_b}{(C_f + C_b) + \dfrac{w}{W}\,(C_b - C_f)} \, < \, \frac{C_f + C_b}{2},
    \label{eq:cp}
\end{equation}
where $w = 4.76$~mm is the channel width and $W$ is the distance between the two outer walls. Equation~(\ref{eq:cp}) follows from interpolating the magnification $M = 1/C$ linearly in depth between the front outer wall ($z = 0$) and the back outer wall ($z = W$), and evaluating it at the front face of the channel, $z = (W - w)/2$, for front and back acrylic walls of equal thickness, which gives $1/C_p = \tfrac{1}{2}(1/C_f + 1/C_b) + \tfrac{w}{2W}(1/C_f - 1/C_b)$. Because the back wall is farther from the lens, $C_b > C_f$, and the interpolated factor is then below the arithmetic mean of the two plate measurements.

\section{Effect of Contrast Enhancement}
\label{app:clahe}
\begin{figure}[h!]
    \centering
    \begin{subfigure}{\linewidth}
        \centering
        \includegraphics[width=0.9\linewidth]{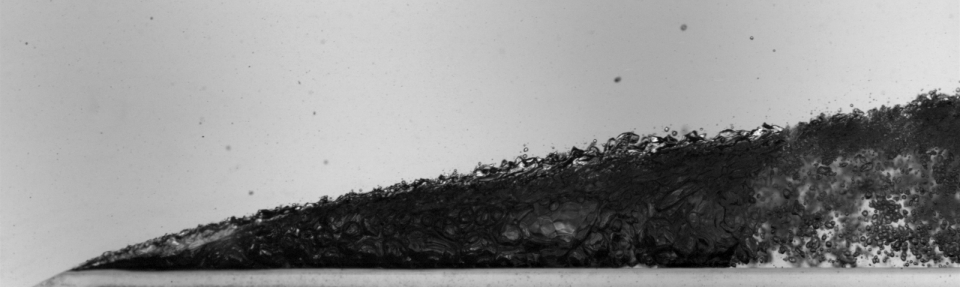}
        \caption{Raw frame}
        \label{fig:clahe_raw}
    \end{subfigure}

    \vspace{0.2cm}
    \begin{subfigure}{\linewidth}
        \centering
        \includegraphics[width=0.9\linewidth]{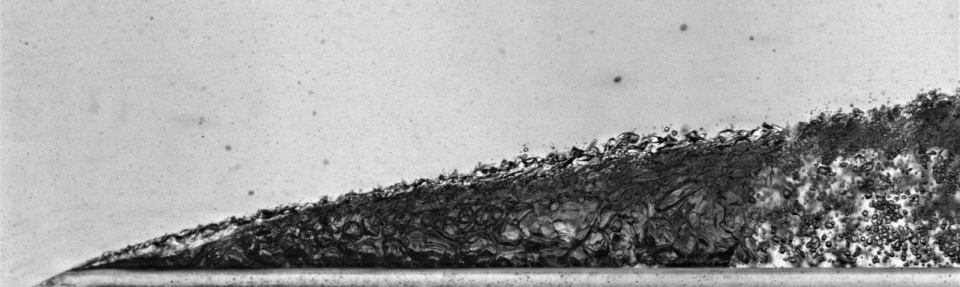}
        \caption{After CLAHE (clip limit $2.0$, $8 \times 8$ tiles)}
        \label{fig:clahe_after}
    \end{subfigure}
    \caption{One recorded frame before and after contrast-limited adaptive histogram equalization. Flow is from left to right.}
    \label{fig:clahe}
\end{figure}

Figure~\ref{fig:clahe} shows one recorded frame before and after the CLAHE step of Section~\ref{sec:algorithms}. In the raw frame the interior of the vapor cloud is compressed into a narrow band of dark gray levels. After equalization the folds and bubble boundaries inside the cavity, the bubbly wake at its closure and the tracer particles in the liquid all stand out from their surroundings. This is the texture the optical flow estimators track, and CLAHE raises it locally without saturating the bright liquid region.

\section{Choice of Pre-trained Checkpoint}
\label{app:pretrained}
\begin{figure}[p]
    \centering
    \begin{subfigure}{\linewidth}
        \centering
        \includegraphics[width=\linewidth]{./figs/01306__truth_field.png}
        \caption{Ground truth from CFD}
        \label{fig:pt_truth}
    \end{subfigure}

    \vspace{0.1cm}
    \begin{subfigure}{\linewidth}
        \centering
        \includegraphics[width=\linewidth]{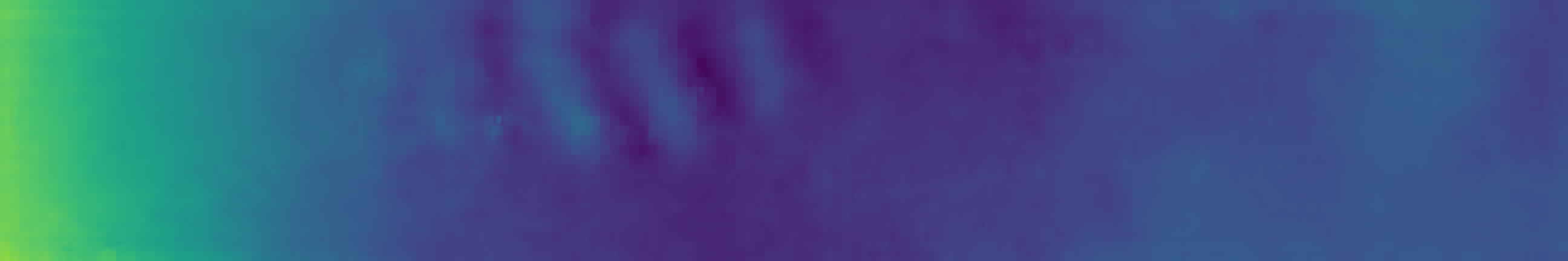}
        \caption{RAFT-Chairs}
        \label{fig:pt_chairs}
    \end{subfigure}

    \vspace{0.1cm}
    \begin{subfigure}{\linewidth}
        \centering
        \includegraphics[width=\linewidth]{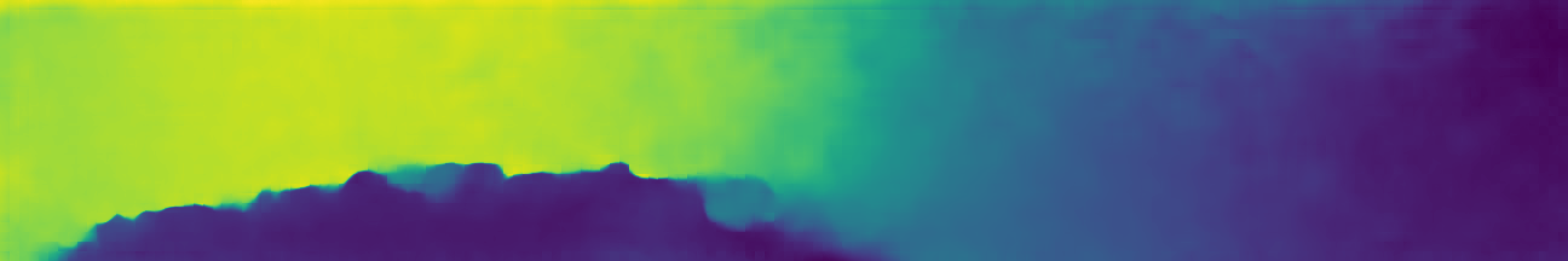}
        \caption{RAFT-KITTI}
        \label{fig:pt_kitti}
    \end{subfigure}

    \vspace{0.1cm}
    \begin{subfigure}{\linewidth}
        \centering
        \includegraphics[width=\linewidth]{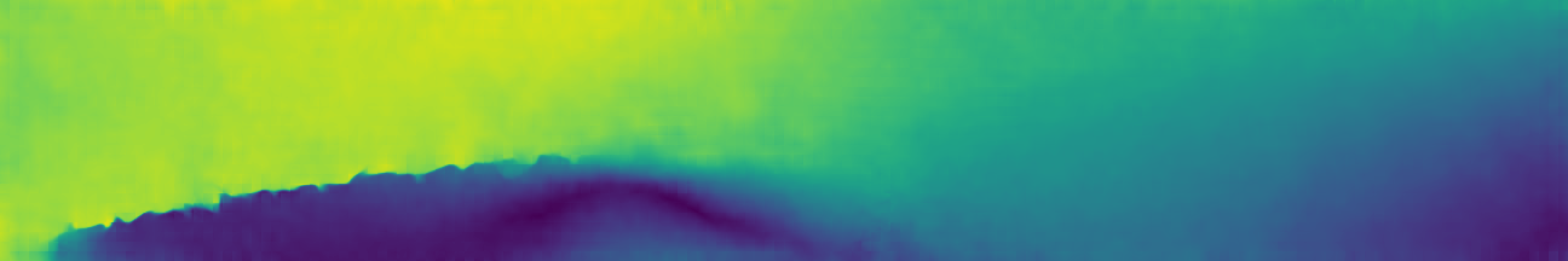}
        \caption{RAFT-Things}
        \label{fig:pt_things}
    \end{subfigure}

    \vspace{0.1cm}
    \begin{subfigure}{\linewidth}
        \centering
        \includegraphics[width=\linewidth]{./figs/01306__raft-sintel_field.png}
        \caption{RAFT-Sintel}
        \label{fig:pt_sintel}
    \end{subfigure}

    \vspace{0.2cm}
    \includegraphics[width=0.5\textwidth]{./figs/cbar_velmag_px.png}
    \caption{Velocity magnitude from the four pre-trained RAFT checkpoints against synthetic ground truth, for the test pair of figure~\ref{fig:velociComparison}. All panels share the color scale shown below them, the velocity magnitude $|\mathbf{u}|$ in px/frame. Flow is from left to right.}
    \label{fig:pretrained}
\end{figure}

RAFT is distributed with weights trained on four generalist datasets, namely FlyingChairs, FlyingThings3D, Sintel and KITTI. Each was run unchanged on the synthetic test pairs before fine-tuning, and the one with the lowest error was used as the starting point. Figure~\ref{fig:pretrained} compares them on the test pair of figure~\ref{fig:velociComparison}, and table~\ref{tab:epe_pretrained} lists the mean end-point errors.

The Sintel checkpoint has the lowest error of the four ($0.300$~px), with FlyingThings3D next ($0.332$~px), while KITTI and FlyingChairs are roughly $1.5$ and $3.2$ times worse. This ordering follows the training data. Sintel contains non-rigid motion, motion blur and atmospheric effects, which are closer to deforming vapor structures than the rigid objects of FlyingChairs or the driving scenes of KITTI. Starting from the Sintel weights, fine-tuning reduced the error by a further factor of two, to $0.152$~px.

\begin{table}[h]
    \centering
    \caption{Mean end-point error of the pre-trained RAFT checkpoints and of the fine-tuned model on the $183$ synthetic test pairs, with an $8$-pixel border left out.}
    \label{tab:epe_pretrained}
    \begin{tabular}{lc}
        \hline
        Model & EPE (px) \\
        \hline
        RAFT-Chairs & $0.953$ \\
        RAFT-KITTI & $0.447$ \\
        RAFT-Things & $0.332$ \\
        RAFT-Sintel & $0.300$ \\
        RAFT-CloudCav (fine-tuned from Sintel) & $0.152$ \\
        \hline
    \end{tabular}
\end{table}

\clearpage
\bibliography{./references}
\bibliographystyle{plainnat}
\end{document}

%% file: tikz/camera_lighting.tex
\begin{tikzpicture}[
    font=\small,
    ar/.style={-{Latex[length=1.8mm]}, thick},
    ray/.style={-{Latex[length=1.6mm]}, orange!80!black, thick},
    scat/.style={-{Latex[length=1.6mm]}, blue!60!black, thick, densely dashed},
]

\begin{scope}
\node[font=\small\bfseries, anchor=west] at (-0.2,4.85) {(i) Top view};

\fill[gray!25] (0,2.05) rectangle (6,2.35);
\fill[cyan!8]  (0,1.0)  rectangle (6,2.05);
\fill[gray!25] (0,0.7)  rectangle (6,1.0);
\draw (0,0.7) rectangle (6,2.35);
\draw (0,1.0) -- (6,1.0);
\draw (0,2.05) -- (6,2.05);
\node[font=\scriptsize, anchor=west] at (0.1,2.2) {far side wall};
\node[font=\scriptsize, anchor=west] at (0.1,0.85) {near side wall};
\draw[ar] (0.4,1.52) -- (1.9,1.52) node[midway, above, font=\scriptsize] {flow};
\draw[{Latex[length=1.3mm]}-{Latex[length=1.3mm]}] (5.6,1.0) -- (5.6,2.05)
      node[midway, left, font=\scriptsize] {$4.76$~mm};

\draw[fill=orange!35] (2.4,3.55) rectangle (3.6,4.15);
\node[font=\scriptsize, anchor=south] at (3.0,4.15) {continuous LED};
\foreach \x in {2.65,3.0,3.35} \draw[ray] (\x,3.55) -- (\x,2.45);

\draw[fill=gray!40] (2.4,-1.35) rectangle (3.6,-0.55);
\draw[fill=gray!55] (2.8,-0.55) -- (3.2,-0.55) -- (3.35,-0.2) -- (2.65,-0.2) -- cycle;
\node[font=\scriptsize, anchor=north] at (3.0,-1.4) {Phantom TMX 6410};
\draw[ar, blue!60!black] (3.0,-0.2) -- (3.0,0.62);
\node[font=\scriptsize, blue!60!black, anchor=west] at (3.1,0.2) {optical axis};
\end{scope}

\begin{scope}[shift={(9.0,0)}]
\node[font=\small\bfseries, anchor=west] at (1.0,4.85) {(ii) Side view};

\fill[gray!25] (0,0.3) rectangle (1.6,2.9);
\fill[cyan!8]  (0.3,0.45) rectangle (1.3,2.75);
\draw (0,0.3) rectangle (1.6,2.9);
\draw (0.3,0.45) rectangle (1.3,2.75);
\node[font=\scriptsize, anchor=east, align=right] at (-0.08,0.85) {far side\\wall};
\node[font=\scriptsize, anchor=west, align=left] at (1.68,2.55) {near side\\wall};
\node[font=\scriptsize, anchor=north, align=center] at (0.8,0.22) {test section\\(cross-section)};
\draw[ar] (-1.75,0.35) -- (-1.75,1.25) node[midway, left, font=\scriptsize] {$y$};

\draw[fill=gray!40] (3.05,1.0) rectangle (3.85,2.2);
\draw[fill=gray!55] (3.05,1.4) -- (3.05,1.8) -- (2.7,1.95) -- (2.7,1.25) -- cycle;
\node[font=\scriptsize, anchor=north] at (3.45,0.95) {camera};
\draw[dashed, blue!60!black] (2.7,1.6) -- (1.6,1.6);

\begin{scope}[shift={(-1.4,3.8)}, rotate=45]
  \draw[fill=orange!35] (-0.6,-0.3) rectangle (0.6,0.3);
\end{scope}
\node[font=\scriptsize, anchor=south] at (-1.4,4.45) {continuous LED};
\draw[ray] (-1.400,3.376) -- (0.338,1.638);
\draw[ray] (-1.188,3.588) -- (0.550,1.850);
\draw[ray] (-0.976,3.800) -- (0.762,2.062);

\draw[dashed, gray] (-0.524,2.5) -- (-1.55,2.5);
\draw[gray] (-1.074,2.5) arc (180:135:0.55);
\node[font=\scriptsize, gray!50!black] at (-1.33,2.82) {$45^\circ$};

\draw[scat] (1.05,1.2) -- (2.6,1.2);
\node[font=\scriptsize, blue!60!black, anchor=north, align=center] at (2.15,1.1) {scattered\\light};
\end{scope}

\end{tikzpicture}

%% file: tikz/raft_architecture.tex
\begin{tikzpicture}[
    font=\small,
    blk/.style={draw, rounded corners=1.5pt, align=center, inner sep=3pt,
                minimum height=11mm, text width=#1},
    img/.style={blk=13mm, fill=gray!12, minimum height=7mm},
    enc/.style={blk=19mm, fill=blue!12},
    corr/.style={blk=21mm, fill=orange!16},
    upd/.style={blk=19mm, fill=green!14},
    flowout/.style={blk=19mm, fill=red!12},
    ar/.style={-{Latex[length=1.8mm]}, thick},
    dar/.style={-{Latex[length=1.8mm]}, thick, dashed},
]

\node[enc]                   (fenc) at (0,0)  {feature\\encoder\\{\scriptsize shared weights\\$1/8$ resolution}};
\node[corr, right=9mm of fenc] (cv)           {4D all-pairs\\correlation volume\\{\scriptsize pyramid $1,2,4,8$}};
\node[corr, right=9mm of cv]  (lk)            {correlation\\lookup\\{\scriptsize around $\mathbf{u}_k$}};
\node[upd,  right=9mm of lk]  (gru)           {ConvGRU\\update\\operator};
\node[flowout, right=9mm of gru] (up)         {convex\\upsampling\\{\scriptsize $\times 8$, learned}};
\node[flowout, right=8mm of up]  (flow)       {$\mathbf{u}=(u,v)$\\{\scriptsize full\\resolution}};

\draw[ar] (fenc) -- (cv);
\draw[ar] (cv)   -- (lk);
\draw[ar] (lk)   -- (gru);
\draw[ar] (gru)  -- (up);
\draw[ar] (up)   -- (flow);

\node[img] (i1) at (-3.6,0.55) {$I_t$};
\node[img] (i2) at (-3.6,-0.55) {$I_{t+1}$};
\node[align=center, font=\scriptsize, below=1mm of i2, text width=26mm]
     {preprocessing\\rotate, crop, CLAHE,\\intensity scaling};
\coordinate (j) at (-2.0,0);
\draw[thick] (i1.east) -- (-2.0,0.55) -- (j);
\draw[thick] (i2.east) -- (-2.0,-0.55) -- (j);
\draw[ar] (j) -- (fenc.west);

\node[enc, above=9mm of gru, yshift=2mm] (cenc) {context encoder\\{\scriptsize $I_t$ only}};
\coordinate (ctop) at ($(cenc.north)+(0,7mm)$);
\draw[thick] (i1.north) -- (i1.north |- ctop) -- (ctop);
\draw[ar] (ctop) -- (cenc.north);
\draw[ar] (cenc) -- (gru);

\coordinate (lo) at ($(gru.south)+(0,-12mm)$);
\coordinate (li) at ($(lk.south)+(0,-12mm)$);
\draw[ar, rounded corners=2pt] (gru.south) -- (lo) -- (li) -- (lk.south);
\node[font=\scriptsize, anchor=north] at ($(lo)!0.5!(li) + (0,-1mm)$)
     {$\mathbf{u}_{k+1}=\mathbf{u}_k+\Delta\mathbf{u}_k$,\ \ $k=1\ldots N$, $N=20$};
\node[font=\scriptsize, anchor=south west] at ($(gru.south east)+(0.5mm,1mm)$) {$\Delta\mathbf{u}_k$};

\node[blk=46mm, fill=yellow!16, below=20mm of cv, xshift=-6mm] (train)
     {\textbf{Training (RAFT-CloudCav)}\\[1pt]
      {\scriptsize weights initialized from the Sintel checkpoint
       and fine-tuned on OpenFOAM-rendered pairs with known $\mathbf{u}_{\text{gt}}$,
       using an $L_1$ sequence loss and AdamW}};
\node[blk=32mm, fill=blue!8, right=16mm of train] (sintel)
     {\textbf{RAFT-Sintel}\\{\scriptsize generalist weights, no fine-tuning (baseline)}};
\draw[dar] (train.east) -- (sintel.west);
\draw[dar] (train.north -| fenc.south) -- (fenc.south)
      node[midway, left, font=\scriptsize] {weights};

\node[blk=40mm, fill=gray!14, below=20mm of up, xshift=14mm] (post)
     {$4\times4$ block averaging $\rightarrow$ HDF5\\
      $\rightarrow$ statistics and uncertainty\\(Section~\ref{sec:uq})};
\draw[ar] (flow.east) -- ++(5mm,0) |- (post.east);

\end{tikzpicture}

%% file: tikz/uq_concept.tex
\begin{tikzpicture}[
    font=\small,
    blk/.style={draw, rounded corners=1.5pt, align=center, inner sep=3pt,
                minimum height=13mm, text width=#1},
    img/.style={blk=13mm, fill=gray!12, minimum height=7mm},
    ar/.style={-{Latex[length=1.8mm]}, thick},
    dar/.style={-{Latex[length=1.8mm]}, thick, dashed},
]

\node[img] (i1) at (-6.2, 0.6) {$I_t$};
\node[img] (i2) at (-6.2,-0.6) {$I_{t+1}$};
\node[blk=17mm, fill=blue!12] (est)  at (-3.9,0) {estimator\\[1pt]$\rightarrow\ \mathbf{u}$};
\node[blk=21mm, fill=green!14] (warp) at (-0.9,0)
     {backward warp\\[1pt]$\tilde{I}_{t+1}=I_{t+1}\!\circ\!\mathbf{u}$};
\node[blk=21mm, fill=orange!16] (ncc) at (2.4,0)
     {NCC surface\\{\scriptsize $I_t$ against $\tilde{I}_{t+1}$,\\local window of\\integer shifts}};

\coordinate (j) at (-5.1,0);
\draw[thick] (i1.east) -- (-5.1,0.6) -- (j);
\draw[thick] (i2.east) -- (-5.1,-0.6) -- (j);
\draw[ar] (j) -- (est.west);
\draw[ar] (est) -- (warp);
\draw[ar] (warp) -- (ncc);

\draw[ar] (i2.south) |- (0,-1.9) -- (warp.south |- 0,-1.9) -- (warp.south);
\draw[ar] (i1.north) |- (0,1.9) -- (ncc.north |- 0,1.9) -- (ncc.north);
\node[font=\scriptsize, anchor=south] at (-1.9,1.95) {ideally $\tilde{I}_{t+1}\equiv I_t$};

\begin{scope}[shift={(6.0,0)}, font=\scriptsize]
  \draw[gray] (-1.5,-1.1) rectangle (1.5,1.1);
  \draw[gray, dashed] (0,-1.1) -- (0,1.1);
  \draw[gray, dashed] (-1.5,0) -- (1.5,0);
  \foreach \r/\o in {0.80/0.58, 0.55/0.40, 0.30/0.22}
     {\draw[orange!70!black] (0.55,0.3) ellipse [x radius=\r, y radius=\o];}
  \fill[red] (0.55,0.3) circle (1.3pt);
  \draw[-{Latex[length=1.6mm]}, red, thick] (0,0) -- (0.55,0.3);
  \node[anchor=south west, red, inner sep=1pt] at (0.18,0.28) {$\boldsymbol{\epsilon}$};
  \node[anchor=north, align=center] at (0,-1.15)
       {NCC surface over integer shifts,\\with its peak $\boldsymbol{\epsilon}$ from zero shift};
\end{scope}
\draw[ar] (ncc.east) -- (4.4,0);

\node[blk=21mm, fill=red!12] (eps) at (6.0,-3.9)
     {residual $\boldsymbol{\epsilon}$ (px)\\{\scriptsize three-point Gaussian sub-pixel fit}};
\node[blk=24mm, fill=yellow!16] (valid) at (10.2,-3.9)
     {validity checks\\{\scriptsize peak outside window, weak correlation, near border $\rightarrow$ masked}};
\node[blk=21mm, fill=gray!14] (sig) at (2.4,-3.9)
     {$\sigma(t)=\langle\|\boldsymbol{\epsilon}\|\rangle_{\text{valid}}$\\{\scriptsize one value per frame pair}};
\node[blk=21mm, fill=gray!14] (sbar) at (-0.9,-3.9)
     {$\sigma_{\bar{u}}=\sqrt{\langle\sigma^2(t)\rangle_t}$\\{\scriptsize equation~(\ref{eq:sigmabar})}};
\node[blk=22mm, fill=red!8] (bands) at (-4.6,-3.9)
     {error bands in figures~\ref{fig:methods48}, \ref{fig:methods30}, \ref{fig:spatial}, \ref{fig:seeding}};

\draw[ar] (6.0,-2.35) -- (eps.north);
\draw[ar] (valid) -- (eps);
\draw[ar] (eps)  -- (sig);
\draw[ar] (sig)  -- (sbar);
\draw[ar] (sbar) -- (bands);

\end{tikzpicture}

%% file: tikz/calibration.tex
\begin{tikzpicture}[
    font=\small,
    dim/.style={{Latex[length=1.4mm]}-{Latex[length=1.4mm]}},
    lab/.style={-{Latex[length=1.6mm]}, gray!60!black},
]
\fill[gray!35] (0,0) rectangle (0.5,4);          
\fill[gray!12] (0.5,0) rectangle (2.0,4);        
\fill[cyan!30] (2.0,0) rectangle (2.6,4);        
\fill[gray!12] (2.6,0) rectangle (4.1,4);        
\fill[gray!35] (4.1,0) rectangle (4.6,4);        
\draw (0,0) rectangle (4.6,4);
\foreach \x in {0.5,2.0,2.6,4.1} \draw (\x,0) -- (\x,4);

\fill[purple!60!black] (-0.12,1.6) rectangle (0,5.0);
\fill[purple!60!black] (4.6,1.6) rectangle (4.72,5.0);
\node[anchor=west, align=left] at (5.0,3.0) {calibration plate\\(front, then back)};
\draw[lab] (5.0,3.1) -- (4.78,4.4);
\draw[lab] (5.0,2.8) .. controls (3.0,2.4) and (0.8,2.6) .. (-0.06,3.6);

\draw[green!50!black, very thick] (2.0,0.8) -- (2.0,5.3);
\node[anchor=east, align=right] at (1.3,5.3) {plane of focus};
\draw[lab] (1.35,5.3) -- (1.95,5.0);

\node[anchor=north] at (0.25,-0.05) {front ($f$)};
\node[anchor=north] at (4.35,-0.05) {back ($b$)};
\draw[dim] (2.0,-0.75) -- (2.6,-0.75) node[midway, below] {$w$};
\draw[dim] (0,-1.45) -- (4.6,-1.45) node[midway, below] {$W$};

\begin{scope}[shift={(-3.4,2.0)}]
  \draw[fill=gray!40] (0,-0.6) rectangle (1.2,0.6);
  \draw[fill=gray!55] (1.2,-0.3) -- (1.6,-0.45) -- (1.6,0.45) -- (1.2,0.3) -- cycle;
  \node[anchor=north] at (0.6,-0.65) {camera};
  \draw[-{Latex[length=1.6mm]}, blue!60!black, thick] (1.6,0) -- (3.2,0);
\end{scope}
\end{tikzpicture}